\documentclass[10pt,sigconf,nonacm]{acmart}

\usepackage{opera}
\usepackage[T1]{fontenc}
\usepackage[utf8]{inputenc}
\usepackage{tabularx}
\usepackage{enumitem}
\usepackage{graphicx}
\usepackage{hyperref}
\usepackage{multirow}
\usepackage{arydshln}
\usepackage{wasysym}
\usepackage{comment}
\usepackage{wrapfig}
\usepackage{subcaption}

\usepackage{tcolorbox}
\tcbuselibrary{skins,breakable}
\definecolor{forestgreen}{rgb}{0.13, 0.54, 0.13}
\newtcolorbox{takeaway}[1]{enforce breakable,enhanced jigsaw,colback=forestgreen!5!white,colframe=forestgreen!75!black,fonttitle=\bfseries,title=#1}

\newcommand{\compz}{compartmentalization~}
\newcommand{\Compz}{Compartmentalization~}
\newcommand{\compzz}{compartmentalization}
\newcommand{\Compzz}{Compartmentalization}

\begin{document}
\def\Snospace~{\S{}}
\renewcommand{\sectionautorefname}{\Snospace}
\renewcommand{\subsectionautorefname}{\Snospace}
\renewcommand{\subsubsectionautorefname}{\Snospace}
\title{Securing Monolithic Kernels using Compartmentalization} %

\author{Soo Yee Lim}
\affiliation{%
	\institution{University of British Columbia}
	\country{Canada}
}

\author{Sidhartha Agrawal}
\affiliation{%
	\institution{University of British Columbia}
	\country{Canada}
}

\author{Xueyuan Han}
\affiliation{%
	\institution{Wake Forest University}
	\country{United States of America}
}

\author{David Eyers}
\affiliation{%
	\institution{University of Otago}
	\country{New Zealand}
}

\author{Dan O'Keeffe}
\affiliation{%
	\institution{Royal Holloway University of London}
	\country{United Kingdom}
}

\author{Thomas Pasquier}
\affiliation{%
	\institution{University of British Columbia}
	\country{Canada}
}

\renewcommand{\shortauthors}{Lim, et al.}



\begin{abstract}
	Monolithic operating systems, where all kernel functionality resides in a single, shared address space, are the foundation of most mainstream computer systems. However, a single flaw, even in a non-essential part of the kernel (\eg device drivers), can cause the entire operating system to fall under an attacker's control. Kernel hardening techniques might prevent certain types of vulnerabilities, but they fail to address a fundamental weakness: the lack of intra-kernel security that safely isolates different parts of the kernel. We survey kernel compartmentalization techniques
that define and enforce intra-kernel boundaries and propose a taxonomy that allows the community to compare and discuss future work. We also identify factors that complicate comparisons among compartmentalized systems, suggest new ways to compare future approaches with existing work meaningfully, and discuss emerging research directions.

\end{abstract}

\maketitle

\section{Introduction}
\label{sec:introduction}
Monolithic operating systems (OSs),
such as Linux, Microsoft Windows,
  and BSD Unix variants, underpin the majority of
  today's desktop, cloud, and mobile computing environments.
While these
OSs support a wide variety of software and hardware,
  their size and complexity leads to frequent,
  severe security flaws~\cite{linuxcves2019}.
  It cannot be assumed
  that these OSs' kernels can be correctly implemented
  to contain security threats from untrusted user-space applications.

Improving the security of monolithic kernels is challenging,
  because any potential solution
  must incur low performance overhead,
  retain compatibility with
  existing applications, and ensure maintainability,
  to be practically adopted.
For example, 
  formally-verified microkernels~\cite{klein2009sel4} 
  and OSs written in memory-safe %
  languages~\cite{redox,cutler2018benefits} offer
  desirable security properties, 
  but they fail to satisfy
  some, if not all, of these requirements.
Consequently,
  their adoption is limited 
  to %
  more niche markets
  (\eg L4 running on Apple's Secure Enclave processors).
On the~other hand,
  incremental approaches to hardening monolithic kernels~\cite{kspp,vander2016protecting}
  are easy to deploy
  but offer more 
  limited security benefits.
For example,
the Kernel Self Protection Project~\cite{kspp}
developed a variety of defenses
for the upstream Linux kernel,
but each eliminates only a specific class of bugs
or a specific method of exploitation.

We discuss \emph{kernel compartmentalization} 
as a way forward
to secure existing monolithic kernel architectures %
with minimal changes to their existing code bases
(and therefore relatively low engineering effort).
Kernel compartmentalization techniques %
divide a monolithic kernel
into multiple compartments,
each within their own security domains,
and provide strong isolation between these domains:
Interactions between domains are restricted
to prevent an attacker from 
taking control of the entire kernel
by exploiting
a vulnerability in just one domain.

After decades of research into kernel compartmentalization,
the time is ripe %
for a systematic look at
past techniques that compartmentalize kernel subsystems
to achieve desirable security properties,
such as confidentiality, integrity, and availability.
Two additional reasons make this survey
particularly timely:
First,
emerging hardware to help support
software compartmentalization
(\eg Arm Morello~\cite{morello})
can substantially reduce performance overhead,
bringing into production
kernel compartmentalization approaches 
that were previously considered too expensive.
Second,
recent interest~\cite{vieira2020fast,cassagnes2020rise,belair2019leveraging,kaffes2021syrup}
in using the extended Berkeley Packet Filter (eBPF)
to import user-supplied code into production kernels at runtime
motivate the development of kernel compartmentalization
to reduce the security risks of running in-kernel eBPF programs.
While eBPF employs a static verifier to guarantee the safety of its programs, 
soundness remains a challenge due to many vulnerabilities in the verifier~\cite{cve-2021-3490,cve-2021-33200,cve-2020-8835,cve-2021-31440,jia2023kernel}.
Kernel compartmentalization could 
enforce runtime isolation of eBPF programs
and therefore allow users to safely extend the kernel~\cite{lim2023ebpf}.

\subsection*{Contributions} 

We synthesize kernel compartmentalization research
spanning over two decades
and make the following contributions:
\vspace{8pt}
\begin{itemize}
	\item \gras{Taxonomy of kernel compartmentalization techniques}:
	We provide a unifying framework 
	to categorize kernel compartmentalization techniques into two broad architectures,
        \emph{sandbox} and \emph{safebox},
        based on their distinct threat models.
	These two architectures differ
        in ways they address
	two important challenges 
	in kernel compartmentalization:
	(1) identifying suitable compartment boundaries 
	and (2) enforcing isolation 
	between the identified boundaries at runtime. 
	\item \gras{Categorization of existing work}:
	We use our taxonomy to discuss the existing body of work 
	and identify potential future research directions.
	\item \gras{Evaluation of kernel compartmentalization}:
	We examine current evaluation strategies 
	used in the literature to understand
	why adopting a general benchmarking strategy
        to compare compartmentalization systems
	is difficult.
	We propose ways to
	simplify future comparisons with existing work.
\end{itemize}

\subsection*{Structure}

In~\autoref{sec:background}, 
we discuss security issues that plague monolithic kernels 
to motivate the need for kernel compartmentalization.
We categorize prior compartmentalization approaches
into two architectures,
\emph{sandbox} and \emph{safebox},
and discuss their differences
in~\autoref{sec:categorisation}.
In~\autoref{sec:boundaries}, 
we analyze
common compartmentalization units
and how their boundaries are identified.
In~\autoref{sec:mechanisms}, 
we classify isolation mechanisms
that enforce these boundaries.
Then,
we examine emerging hardware technologies
that will likely 
influence future compartmentalization landscape 
in~\autoref{sec:emerging_hw}.
Finally, 
in \autoref{sec:evaluation}, 
we discuss the limitations of the
evaluation strategies currently used in the literature 
and propose ways to better
understand and compare different compartmentalization solutions.

\section{Background and Motivation}
\label{sec:background}
\begin{figure*}
        \includegraphics[width=\textwidth]{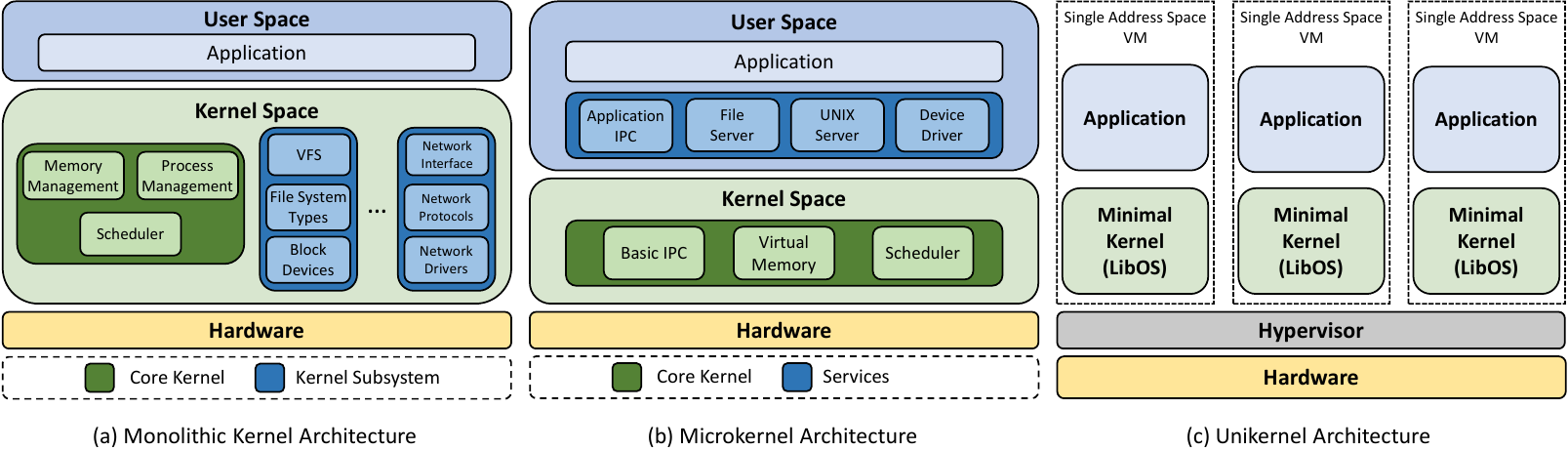}
        \caption{Alternative kernel designs}
	\label{fig:kernel_architectures}
\end{figure*}

A monolithic OS design (\autoref{fig:kernel_architectures}a)
is problematic with respect to security
mainly for two reasons.
First,
the size of a monolithic kernel's code base
and its complexity
make it easy to introduce
bugs and vulnerabilities,
which could allow untrusted user-space applications 
to compromise the kernel and thus the entire system.
For example, %
the \texttt{SLOCCount} tool~\cite{sloccount}, 
which we use to measure the size of the Linux v6.4 kernel code base,
reported $\sim$24 million lines of code,
98\% of which is written in memory-unsafe C
and 1.2\% in assembly. %
While individual OS instances typically
use only a subset of this large code base, 
their sizes are still substantial
compared to those of the vast majority of software.
With a steady development rate of over 75,000 commits per year~\cite{2020linuxreport},
more than 1,000 vulnerabilities were discovered in the last five years~\cite{linux_vulntrend}.
Applying formal verification~\cite{klein2009operating} and static program
analysis~\cite{gens2018k} to soundly prove
the absence of bugs and vulnerabilities in the code base
is difficult,
because existing %
approaches
do not scale to the size of the kernel.
For instance, the seL4 project took 11 person-years 
to formally verify a microkernel that has about 9,000 lines of code~\cite{klein2009sel4}.
Similarly, 
recent kernel fuzzing approaches~\cite{schumilo2017kafl,jeong2019razzer,kim2020hfl,song2020agamotto} 
can detect 
previously unknown vulnerabilities,
but they cannot \emph{guarantee} 
the absence of vulnerabilities in the kernel.
Finally,
rewriting existing kernel code 
in a safer language 
(\eg Rust for Linux~\cite{rust-kernel}) 
can reduce the prevalence of 
some (\eg out-of-bounds array access)
but not all types of vulnerabilities.
For instance, %
Rust cannot prevent memory leaks
that lead to resource exhaustion~\cite{klabnik2023rust}.

Second, 
all OS services, including
the core functionality (scheduling, memory management, \etc),
modules that extend the kernel (\eg audit systems),
and device drivers,
operate in a single security domain.
This lack of isolation
can give an attacker %
full privilege to access,
modify, and control the entire
system's data and resources
when they exploit a vulnerability
\emph{anywhere} in the kernel.
Even a bug in a peripheral kernel module 
can constitute a single point of failure,
causing essential OS functionality
to fall fully under the attacker's control.
For example,
a heap-based buffer overflow vulnerability (CVE-2020-12654) 
in the Linux kernel's Marvell Wi-Fi driver
can cause arbitrary code execution and denial-of-service (DoS)~\cite{cve2020_12654}.

\subsection{Kernel Hardening Techniques}

One strategy to address the fundamental brittleness of 
a monolithic kernel architecture
is to increase the barrier to an attack.
For example, 
address space layout randomization (ASLR)~\cite{giuffrida2012enhanced} 
randomizes the locations of libraries,
the heap, and the stack 
in a process' address space 
so that
an attacker cannot reliably
find useful instructions to create 
a return-oriented programming attack~\cite{giuffrida2012enhanced}.
Other techniques to make bugs harder to exploit 
include data execution prevention (DEP)~\cite{andersen2004data},
stack canaries~\cite{wagle2003stackguard},
and shadow stacks~\cite{chiueh2001rad, davi2011ropdefender, jie2020silhouette}.
All of these techniques, 
commonly referred to as \emph{kernel hardening},
are effective at reducing the possibility 
of a vulnerability being successfully exploited,
but they can also be bypassed by attackers~\cite{snow2013just, hund2013practical, gras2017aslr, evtyushkin2016jump, stojanovski2007bypassing, bierbaumer2018smashing, lan2015loop}.
For example,
attackers can exploit information leakage
from side channels
to deduce~\cite{barresi2015cain}
and even derandomize~\cite{gadient2019automatic} 
address space layout.
Kernel compartmentalization is complementary 
to existing kernel hardening techniques
and can be deployed alongside them.

\subsection{Alternative Kernel Designs}
\label{sec:background:alternatives}
Microkernels and unikernels
are the two main alternatives
to monolithic kernels, 
see~\autoref{fig:kernel_architectures}b 
and~\autoref{fig:kernel_architectures}c.
We briefly discuss them here
and refer the interested reader
to existing literature~\cite{l3tosel4, madhavapeddy2014unikernels}
for more detailed discussions and
different interpretations
of these designs.
Other kernel designs (\eg exokernel~\cite{engler1995exokernel} and multikernel~\cite{baumann2009multikernel})  exist,       
but since we focus on monolithic kernels, 
an in-depth discussion 
of these designs is beyond the scope of this survey.

\noindgras{Microkernel:} 
Unlike a monolithic kernel,
a microkernel~\cite{l3tosel4} implements
only a %
subset
of kernel functionality
(\eg scheduling and %
IPC),
moving other %
functionality,
such as file systems and device drivers,
into user space.
This \emph{compartmentalized} design
(where individual subsystems
run in their own, mostly isolated, domains)
significantly reduces the system's 
trusted computing base (TCB)
to the point where 
the kernel code can be formally verified~\cite{klein2009sel4}.
However,
microkernels can incur large
performance overhead,
because switching between the kernel
and user-mode functionality is costly,
especially for high-throughput devices
such as network cards.
As we see in~\autoref{sec:mechanisms},
some kernel compartmentalization systems
use similar mechanisms
to separate functionality
in a monolithic kernel,
inheriting similar performance~issues.

\noindgras{Unikernel:}
A unikernel (and a library OS in general)
builds a single address space OS stack
that only contains the OS subsystems required by
a specific application.
Therefore, 
compared to a microkernel, %
a unikernel does not incur the expensive cost of
context switches;
compared to a classic monolithic kernel,
it reduces the TCB
by excluding unnecessary kernel functionality.
However,
since all kernel code essential to a~target application
shares the same address space,
a vulnerability in one kernel component
(\eg a buggy device driver)
can still affect the entire kernel stack.
Consequently, 
some unikernels~\cite{lefeuvre2021flexos,sartakov21} resort to
compartmentalization techniques.
Moreover,
while recent unikernels
(\eg Lupine OS~\cite{kuo2020linux} and Unikraft~\cite{unikraft})
can run legacy programs,
unlike a monolithic kernel,
their highly specialized,
application-specific organization
is unsuitable for general-purpose computing.

\noindgras{The future of monolithic kernels: }
Alternatives to monolithic kernel designs 
are gradually conquering a wider portion of the global OS market (\eg MINIX~\cite{herder2006minix} on Intel chips, 
and embedded L4~\cite{l3tosel4} on Apple devices).
However, a significantly large proportion of devices still depend on monolithic OSs 
and will continue to do so in the foreseeable future.
We believe that compartmentalization of a monolithic OS
is essential to an \emph{incremental} path
towards a safer OS~\cite{li2021incremental}. %
Anecdotally, while not strictly kernel compartmentalization, 
the Linux community has shown interest in address space isolation techniques 
to combat the Spectre/Meltdown family of vulnerabilities~\cite{asi_lwn}.
Some compartmentalization systems,
such as VirtuOS~\cite{nikolaev_virtuos_2013}, 
have gone as far as effectively transforming monolithic OSs into microkernels,
blurring the line between these two architectures.

\subsection{Software Compartmentalization}
\label{sec:background:software}

Kernel compartmentalization
is a type of \emph{software compartmentalization},
which is more generally defined as
the process of decomposing a software artifact
into isolated but collaborative components
to mitigate the exploitation of vulnerabilities~\cite{gudka2015clean}.
Many software compartmentalization techniques
have been proposed in the user-space context, which generally fall into three categories:
(1) protecting the OS from untrusted \emph{user-space} applications~\cite{douceur2008leveraging,yee2010native,jana2011txbox,kim2013practical},
(2) protecting security-sensitive user-space applications or pieces of application logic (PAL) from a compromised OS~\cite{garfinkel2003terra,mccune2008flicker,mccune2010trustvisor,steinberg2010nova,keller2010nohype,azab2011sice,hofmann2013inktag}, and
(3) protecting \emph{both} the OS from misbehaving user-space applications and user-space applications from a malicious OS~\cite{li2014minibox}.
The interested reader can refer to the survey by Shu et al.~\cite{shu2016study}.

While we see similar use cases
emerge in kernel space (\autoref{sec:categorisation}),
we focus on two important aspects of software compartmentalization
in the kernel context:
(1) identifying compartment boundaries (\autoref{sec:boundaries}),
and (2) enforcing these boundaries
while maintaining good performance (\autoref{sec:mechanisms}).
Differences between the techniques employed in user space 
and within the kernel
are significant enough that
a direct application from one domain
to the other is challenging (see~\autoref{sec:boundaries:decomposition}).

\section{Overview of Kernel \Compz Approaches}
\label{sec:categorisation}
\begin{table*}[t]
\centering
\caption{An overview of existing kernel \compz systems (M = Manual, A = Automated, N/R = Not Required)}
\label{tab:decomposition}
\resizebox{\textwidth}{!}{%
\begin{tabular}{ll|l||l||cccl||cc||lclll|}
\cline{3-15}
 &
   &
  \multicolumn{1}{c||}{\multirow{11}{*}{\textbf{Use Case}}} &
  \multicolumn{1}{c||}{\multirow{11}{*}{\textbf{Paper}}} &
  \multicolumn{4}{c||}{\multirow{3}{*}{\textbf{Security Properties}}} &
  \multicolumn{2}{c||}{\multirow{3}{*}{\textbf{Boundary}}} &
  \multicolumn{5}{c|}{\textbf{Isolation Mechanism}} \\ \cline{11-15} 
 &
   &
  \multicolumn{1}{c||}{} &
  \multicolumn{1}{c||}{} &
  \multicolumn{4}{c||}{} &
  \multicolumn{2}{c||}{} &
  \multicolumn{3}{c|}{\textbf{Intra-kernel}} &
  \multicolumn{1}{l|}{\multirow{10}{*}{\textbf{User/Kernel}}} &
  \multirow{10}{*}{\textbf{Kernel/Hypervisor}} \\ \cline{11-13}
\multicolumn{1}{c}{\textbf{}} &
  \multicolumn{1}{c|}{\textbf{}} &
  \multicolumn{1}{c||}{} &
  \multicolumn{1}{c||}{} &
  \multicolumn{4}{c||}{} &
  \multicolumn{2}{c||}{} &
  \multicolumn{2}{c|}{\textbf{Domain Based}} &
  \multicolumn{1}{c|}{\multirow{8}{*}{\textbf{\rotatebox{90}{Address Based}}}} &
  \multicolumn{1}{l|}{} &
   \\ \cline{5-12}
\textbf{} &
  \textbf{} &
  \multicolumn{1}{c||}{} &
  \multicolumn{1}{c||}{} &
  \multicolumn{1}{c|}{\textbf{\rotatebox{90}{Confidentiality}}} &
  \multicolumn{1}{c|}{\textbf{\rotatebox{90}{Integrity}}} &
  \multicolumn{1}{c|}{\textbf{\rotatebox{90}{Reliability}}} &
  \multicolumn{1}{c||}{\textbf{\rotatebox{90}{Recoverability}}} &
  \multicolumn{1}{c|}{\textbf{\rotatebox{90}{Decomposition}}} &
  \multicolumn{1}{c||}{\textbf{\rotatebox{90}{Glue Code Generation\hspace{0.5em}}}} &
  \multicolumn{1}{c|}{\textbf{\rotatebox{90}{Software Based}}} &
  \multicolumn{1}{c|}{\textbf{\rotatebox{90}{Hardware Assisted}}} &
  \multicolumn{1}{c|}{} &
  \multicolumn{1}{l|}{} &
   \\ \hline
\multicolumn{1}{|l|}{\multirow{18}{*}{\rotatebox[origin=c]{90}{\textbf{Architecture}}}} &
  \multirow{12}{*}{\rotatebox[origin=c]{90}{\textbf{Sandbox}}} &
  \multirow{3}{*}{Fault Isolation} &
  Mondrix~\cite{witchel2005mondrix} &
  \multicolumn{1}{c|}{} &
  \multicolumn{1}{c|}{$\bullet$} &
  \multicolumn{1}{c|}{$\bullet$} &
   &
  \multicolumn{1}{c|}{M} &
  M &
  \multicolumn{1}{l|}{} &
  \multicolumn{1}{c|}{$\bullet$} &
  \multicolumn{1}{l|}{} &
  \multicolumn{1}{l|}{} &
   \\ \cline{4-15} 
\multicolumn{1}{|l|}{} &
   &
   &
  Microdrivers~\cite{ganapathy_microdrivers_2008} &
  \multicolumn{1}{c|}{} &
  \multicolumn{1}{c|}{$\bullet$} &
  \multicolumn{1}{c|}{$\bullet$} &
   &
  \multicolumn{1}{c|}{A} &
  A &
  \multicolumn{1}{l|}{} &
  \multicolumn{1}{l|}{} &
  \multicolumn{1}{c|}{} &
  \multicolumn{1}{c|}{$\bullet$} &
   \\ \cline{4-15} 
\multicolumn{1}{|l|}{} &
   &
   &
  Decaf~\cite{renzelmanndecaf} &
  \multicolumn{1}{c|}{} &
  \multicolumn{1}{c|}{$\bullet$} &
  \multicolumn{1}{c|}{$\bullet$} &
   &
  \multicolumn{1}{c|}{A} &
  A &
  \multicolumn{1}{l|}{} &
  \multicolumn{1}{l|}{} &
  \multicolumn{1}{c|}{} &
  \multicolumn{1}{c|}{$\bullet$} &
   \\ \cline{3-15} 
\multicolumn{1}{|l|}{} &
   &
  \multirow{4}{*}{Fault Resistance} &
  Nooks~\cite{swift2003improving} &
  \multicolumn{1}{c|}{} &
  \multicolumn{1}{c|}{$\bullet$} &
  \multicolumn{1}{c|}{$\bullet$} &
  \multicolumn{1}{c||}{$\bullet$} &
  \multicolumn{1}{c|}{M} &
  M &
  \multicolumn{1}{l|}{} &
  \multicolumn{1}{l|}{} &
  \multicolumn{1}{c|}{$\bullet$} &
  \multicolumn{1}{c|}{} &
   \\ \cline{4-15} 
\multicolumn{1}{|l|}{} &
   &
   &
  BGI~\cite{castro2009fast} &
  \multicolumn{1}{c|}{} &
  \multicolumn{1}{c|}{$\bullet$} &
  \multicolumn{1}{c|}{$\bullet$} &
  \multicolumn{1}{c||}{$\bullet$} &
  \multicolumn{1}{c|}{M} &
  N/R &
  \multicolumn{1}{c|}{$\bullet$} &
  \multicolumn{1}{c|}{} &
  \multicolumn{1}{l|}{} &
  \multicolumn{1}{c|}{} &
   \\ \cline{4-15} 
\multicolumn{1}{|l|}{} &
   &
   &
  SafeDrive~\cite{safedrive06} &
  \multicolumn{1}{l|}{} &
  \multicolumn{1}{c|}{$\bullet$} &
  \multicolumn{1}{c|}{$\bullet$} &
  \multicolumn{1}{c||}{$\bullet$} &
  \multicolumn{1}{c|}{M} &
  N/R &
  \multicolumn{1}{c|}{$\bullet$} &
  \multicolumn{1}{c|}{} &
  \multicolumn{1}{l|}{} &
  \multicolumn{1}{l|}{} &
   \\ \cline{4-15} 
\multicolumn{1}{|l|}{} &
   &
   &
  VirtuOS~\cite{nikolaev_virtuos_2013} &
  \multicolumn{1}{c|}{} &
  \multicolumn{1}{c|}{$\bullet$} &
  \multicolumn{1}{c|}{$\bullet$} &
  \multicolumn{1}{c||}{$\bullet$} &
  \multicolumn{1}{c|}{M} &
  M &
  \multicolumn{1}{l|}{} &
  \multicolumn{1}{l|}{} &
  \multicolumn{1}{l|}{} &
  \multicolumn{1}{l|}{} &
  \multicolumn{1}{c|}{$\bullet$} \\ \cline{3-15} 
\multicolumn{1}{|l|}{} &
   &
  \multirow{5}{*}{Vulnerability Isolation} &
  XFI~\cite{erlingsson2006xfi} &
  \multicolumn{1}{c|}{} &
  \multicolumn{1}{c|}{$\bullet$} &
  \multicolumn{1}{c|}{} &
   &
  \multicolumn{1}{c|}{A} &
  N/R &
  \multicolumn{1}{c|}{$\bullet$} &
  \multicolumn{1}{c|}{} &
  \multicolumn{1}{l|}{} &
  \multicolumn{1}{c|}{} &
   \\ \cline{4-15} 
\multicolumn{1}{|l|}{} &
   &
   &
  LXFI~\cite{mao_software_2011} &
  \multicolumn{1}{c|}{} &
  \multicolumn{1}{c|}{$\bullet$} &
  \multicolumn{1}{c|}{} &
   &
  \multicolumn{1}{c|}{M} &
  N/R &
  \multicolumn{1}{c|}{$\bullet$} &
  \multicolumn{1}{c|}{} &
  \multicolumn{1}{l|}{} &
  \multicolumn{1}{c|}{} &
   \\ \cline{4-15} 
\multicolumn{1}{|l|}{} &
   &
   &
  LXDs~\cite{Narayanan2019} &
  \multicolumn{1}{c|}{$\bullet$} &
  \multicolumn{1}{c|}{$\bullet$} &
  \multicolumn{1}{c|}{} &
   &
  \multicolumn{1}{c|}{M} &
  A &
  \multicolumn{1}{l|}{} &
  \multicolumn{1}{l|}{} &
  \multicolumn{1}{l|}{} &
  \multicolumn{1}{l|}{} &
  \multicolumn{1}{c|}{$\bullet$} \\ \cline{4-15} 
\multicolumn{1}{|l|}{} &
   &
   &
  LVD~\cite{narayanan2020lightweight} &
  \multicolumn{1}{c|}{$\bullet$} &
  \multicolumn{1}{c|}{$\bullet$} &
  \multicolumn{1}{c|}{} &
   &
  \multicolumn{1}{c|}{M} &
  A &
  \multicolumn{1}{l|}{} &
  \multicolumn{1}{l|}{} &
  \multicolumn{1}{l|}{} &
  \multicolumn{1}{l|}{} &
  \multicolumn{1}{c|}{$\bullet$} \\ \cline{4-15} 
\multicolumn{1}{|l|}{} &
   &
   &
  HAKC~\cite{mckee2022preventing} &
  \multicolumn{1}{c|}{$\bullet$} &
  \multicolumn{1}{c|}{$\bullet$} &
  \multicolumn{1}{c|}{} &
  \multicolumn{1}{c||}{} &
  \multicolumn{1}{c|}{M} &
  M &
  \multicolumn{1}{c|}{} &
  \multicolumn{1}{c|}{$\bullet$} &
  \multicolumn{1}{c|}{} &
  \multicolumn{1}{c|}{} &
  \multicolumn{1}{c|}{} \\ \cline{2-15} 
\multicolumn{1}{|l|}{} &
  \multicolumn{1}{c|}{\multirow{6}{*}{\rotatebox[origin=c]{90}{\textbf{Safebox}}}} &
  \multirow{4}{*}{Security State Isolation} &
  KCoFI~\cite{criswell2014kcofi} &
  \multicolumn{1}{c|}{} &
  \multicolumn{1}{c|}{$\bullet$} &
  \multicolumn{1}{c|}{} &
   &
  \multicolumn{1}{c|}{M} &
  N/R &
  \multicolumn{1}{l|}{} &
  \multicolumn{1}{c|}{$\bullet$} &
  \multicolumn{1}{c|}{} &
  \multicolumn{1}{c|}{} &
   \\ \cline{4-15} 
\multicolumn{1}{|l|}{} &
  \multicolumn{1}{c|}{} &
   &
  KENALI~\cite{song2016enforcing} &
  \multicolumn{1}{c|}{} &
  \multicolumn{1}{c|}{$\bullet$} &
  \multicolumn{1}{c|}{} &
   &
  \multicolumn{1}{c|}{A} &
  N/R &
  \multicolumn{1}{l|}{} &
  \multicolumn{1}{c|}{$\bullet$} &
  \multicolumn{1}{c|}{} &
  \multicolumn{1}{c|}{} &
   \\ \cline{4-15} 
\multicolumn{1}{|l|}{} &
  \multicolumn{1}{c|}{} &
   &
  xMP~\cite{proskurin2020xmp} &
  \multicolumn{1}{c|}{$\bullet$} &
  \multicolumn{1}{c|}{$\bullet$} &
  \multicolumn{1}{c|}{} &
   &
  \multicolumn{1}{c|}{M} &
  N/R &
  \multicolumn{1}{l|}{} &
  \multicolumn{1}{l|}{} &
  \multicolumn{1}{l|}{} &
  \multicolumn{1}{l|}{} &
  \multicolumn{1}{c|}{$\bullet$} \\ \cline{4-15} 
\multicolumn{1}{|l|}{} &
  \multicolumn{1}{c|}{} &
   &
  IskiOS~\cite{iskios_raid_21} &
  \multicolumn{1}{c|}{$\bullet$} &
  \multicolumn{1}{c|}{$\bullet$} &
  \multicolumn{1}{l|}{} &
   &
  \multicolumn{1}{c|}{M} &
  N/R &
  \multicolumn{1}{c|}{} &
  \multicolumn{1}{c|}{$\bullet$} &
  \multicolumn{1}{l|}{} &
  \multicolumn{1}{l|}{} &
   \\ \cline{3-15} 
\multicolumn{1}{|l|}{} &
  \multicolumn{1}{c|}{} &
  \multirow{2}{*}{Secure Policy Enforcement} &
  PerspicuOS~\cite{dautenhahn2015nested} &
  \multicolumn{1}{c|}{} &
  \multicolumn{1}{c|}{$\bullet$} &
  \multicolumn{1}{c|}{} &
   &
  \multicolumn{1}{c|}{N/R} &
  N/R &
  \multicolumn{1}{l|}{} &
  \multicolumn{1}{l|}{} &
  \multicolumn{1}{c|}{$\bullet$} &
  \multicolumn{1}{c|}{} &
   \\ \cline{4-15} 
\multicolumn{1}{|l|}{} &
  \multicolumn{1}{c|}{} &
   &
  SKEE~\cite{azab_skee_2016} &
  \multicolumn{1}{c|}{} &
  \multicolumn{1}{c|}{$\bullet$} &
  \multicolumn{1}{c|}{} &
   &
  \multicolumn{1}{c|}{N/R} &
  N/R &
  \multicolumn{1}{l|}{} &
  \multicolumn{1}{l|}{} &
  \multicolumn{1}{c|}{$\bullet$} &
  \multicolumn{1}{c|}{} &
   \\ \hline
\end{tabular}%
}
\end{table*}

\begin{figure}
	\centering
	\includegraphics[width=\columnwidth]{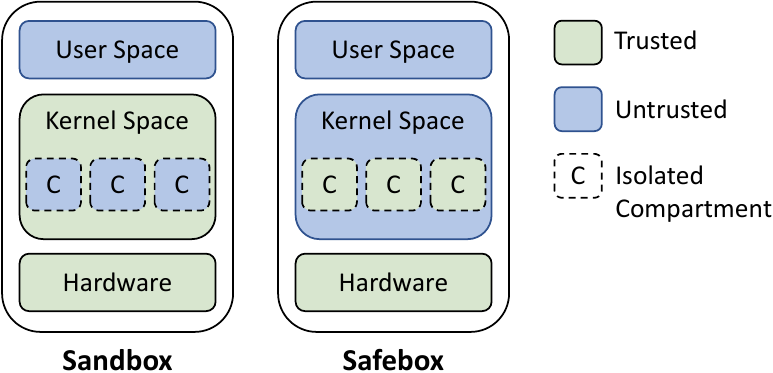}
	\caption{Threat models of kernel \compz architectures.}
	\label{img:threatmodel}
\end{figure}

Generally speaking,
all kernel compartmentalization systems
create boundaries between
\emph{untrusted} system components,
which might be buggy or malicious,
and \emph{trusted} ones, which are assumed to be benign.
Kernel compartmentalization is enforced by
(1) \emph{resource access control}
to restrict the code and data
accessible to a kernel compartment,
and (2) \emph{control transfer restriction}
when execution crosses a compartment boundary.
However,
existing compartmentalization systems differ
in the \emph{direction}
in which
these controls are performed
at the compartment boundary
due to different assumptions
of the threat model they adopt (\autoref{img:threatmodel}).
We categorize
existing kernel compartmentalization work
into the \emph{sandbox} architecture
and the \emph{safebox} architecture (\autoref{tab:decomposition}).

\noindgras{Sandbox}:
Some kernel compartmentalization systems~\cite{castro2009fast,ganapathy_microdrivers_2008,renzelmanndecaf}
target kernel subsystems, such as device drivers,
that traditionally run in kernel space
but are typically considered to be less trustworthy
than the rest of the kernel.
These compartmentalization systems enforce
security rules on data and control flow
going \emph{out of} the compartment boundary,
so that any bugs or vulnerabilities within the compartment
cannot adversely affect the kernel.
The sandbox architecture reduces \emph{the kernel's attack surface}
by confining an untrustworthy component within its own isolated compartments.

\noindgras{Safebox}:
Safebox systems~\cite{dautenhahn2015nested,azab_skee_2016}
follow the opposite threat model.
They assume that the kernel is untrustworthy
and therefore isolate some critical kernel functionalities,
such as mandatory access control, %
to protect them
from potential threats originating from the rest of the kernel.
These systems enforce
security rules on data and control flow
going \emph{into} the compartment boundary.
From the point of view of an isolated component,
the safebox architecture reduces \emph{the component's attack surface}
by excluding the rest of the kernel from its TCB.

We discuss both architectures in more detail
in the remainder of this section,
identifying the security properties
that each architecture serves,
for various use cases.

\subsection{Sandbox Architecture}
\label{sec:categorisation:exclave}
Kernel extensions, such as loadable kernel modules and device drivers, have long been a major source of errors~\cite{chou2001empirical} and vulnerabilities~\cite{chen2011linux,palix2011faults,linux_vulntrend} in the Linux kernel.
The sandbox architecture
protects the kernel
from their errors or exploits
by isolating each of these components in
a separate compartment
while restricting resource accesses and control transfers
\emph{out of} the compartment's boundary.

\noindgras{Resource Access Control.}
Resource access control preserves the integrity
of the kernel's \emph{non-control data}
(\eg configuration data)
by restricting write access
to the data
outside of an isolated compartment.
This protects the rest~of~the kernel
from any modification
that could corrupt kernel memory
and change the kernel's behavior unexpectedly.

\noindgras{Control Transfer Restriction.}
The corruption of \emph{control data}
(\eg function pointers, jump targets, and return addresses)
diverts a program's normal control flow.
To prevent an isolated component
from executing arbitrary kernel code
in the event of an error or kernel exploit,
sandbox systems interpose on all control transfers
\emph{out of} the compartment's boundary.
Specifically,
they restrict the component's control flow
to an immutable set of valid entry
points,
which can be defined
at compile time \eg
as %
an allowlist
of control flow transfer targets~\cite{swift2003improving,castro2009fast,erlingsson2006xfi}.

\noindgras{Use Cases and Security Properties.}
The literature covers three main types of threat models
regarding kernel extensions against which the sandbox architecture protects the core kernel:
(1) \emph{buggy} extensions containing bugs that could affect system up-time;
(2) \emph{exploitable} extensions containing bugs that can be leveraged
by a user-space program to corrupt kernel state (\eg privilege escalation);
(3) \emph{malicious} extensions actively trying to compromise the system
(\eg kernel rootkits).
These models represent different real-life attack scenarios.
For example,
the \emph{buggy} threat model is suitable for an enterprise workstation
where employees have no root access, and kernel extensions are centrally managed.
A user under the \emph{exploitable} threat model
might unintentionally install a piece of untrusted software
that exploits bugs in kernel extensions.
Instead,
under the \emph{malicious} threat model,
an insider or supply-chain attack
might be used to install a kernel rootkit to infiltrate the victim machine.
Sandbox systems confine \emph{buggy} extensions through \emph{fault isolation}
and \emph{fault resistance},
and contain \emph{exploitable} and \emph{malicious} extensions
via \emph{vulnerability isolation}.
Overall, they achieve \emph{integrity},
\emph{confidentiality},
\emph{reliability},
and \emph{recoverability}.

\subsubsection{Fault Isolation}
\label{sec:categorisation:exclave:faultisolation}
Faults in the kernel could corrupt and crash the system.
Unfortunately,
they often arise due to bugs %
in the large and complex kernel code base.
Fault isolation~\cite{witchel2005mondrix,ganapathy_microdrivers_2008,renzelmanndecaf, mckee2022preventing} confines the impact of an error
in the buggy
component
to improve
a system's
\emph{integrity} and \emph{reliability}.

\noindent{\bf{\emph{Integrity.}}}
Commodity monolithic kernels are implemented in memory unsafe languages
and thus are vulnerable to memory corruption.
A bug in a kernel component can
e.g., corrupt critical kernel state
that leads to system failure
or otherwise unexpected behavioral changes
in the system~\cite{Biba1977}.
Fault-isolating systems restrict write access to kernel memory
to ensure that a sandboxed kernel component can modify only
its own memory or the memory it has legitimate access to.
This preserves \emph{data integrity},
which prevents an error
from modifying kernel memory outside the sandbox.
Additionally,
a sandboxed component cannot execute arbitrary %
kernel functions without having legitimate access to them.
This restriction enforces \emph{control-flow integrity},
which, combined with resource access control
(i.e., data integrity),
prevents the buggy kernel component
from affecting the behavior of the rest of the system.

\noindent{\bf{\emph{Reliability.}}}
Reliability measures a system's ability
to remain operational
despite the presence of misbehaving components.
By compartmentalizing unreliable kernel components
and restricting their interactions
with the rest of the system,
existing kernel compartmentalization techniques~\cite{swift2003improving,castro2009fast,nikolaev_virtuos_2013,witchel2005mondrix,ganapathy_microdrivers_2008,renzelmanndecaf,safedrive06}
allow faulty components to fail without crashing the system,
improving its reliability.
The end goal is to increase overall system up-time.

\subsubsection{Fault Resistance}
While a fault-isolating system
contains the impact of an error,
a faulty component remains unavailable
until it is reloaded
or when the system reboots.
To restore (some of) the component's functionality,
a fault-resistant system~\cite{swift_recovering_2004,castro2009fast,nikolaev_virtuos_2013,safedrive06}
not only isolates
but also automatically recovers from most
(but not necessarily all)
faults within a compartment.
Therefore,
in addition to a fault-isolation mechanism,
a fault-resistant system also implements
a recovery mechanism.
For example,
a virtual machine (VM) based compartmentalization system~\cite{nikolaev_virtuos_2013}
can leverage the hypervisor's ability
to terminate and restart VMs
to recover from a failed compartment.
Thus,
such a system also guarantees
\emph{recoverability}.

\noindent{\bf{\emph{Recoverability.}}}
A kernel component's failure
affects
the availability
of its dependent components.
To facilitate automatic recovery, a fault-resistant system tracks
the component's updates to kernel objects.
When a failure is detected, the recovery mechanism unloads the faulty component, restores kernel state, releases kernel resources held by the faulty component, and restarts the kernel component in its compartment.
This restores the faulty component to a functional state from which it can continue;
as a result, other affected components can return to normal operations.

\subsubsection{Vulnerability Isolation}
Both fault-isolating and fault-resistant systems
assume that faults in a kernel component are unintended errors.
In contrast, vulnerability-isolating systems~\cite{erlingsson2006xfi,mao_software_2011,Narayanan2019,narayanan2020lightweight} assume a stronger threat model where the component is either exploitable or malicious.
Since they target kernel attacks rather than faults,
they provide no reliability or recoverability guarantees;
instead, they guarantee \emph{confidentiality} and \emph{integrity}.

\noindent{\bf{\emph{Confidentiality.}}}
An adversary with an arbitrary read primitive
can access any sensitive information in kernel memory.
For example,
a vulnerability~\cite{cve-2018-15471}
in the Xen network back-end driver
allows for out-of-bounds access,
which leaks kernel information.
A sandbox system can restrict
an untrustworthy component's read access
to kernel memory outside of its compartment
to protect sensitive kernel data.
However, software-based vulnerability-isolating systems generally
do not restrict read access for performance concerns;
consequently, confidentiality attacks are out of their scope~\cite{erlingsson2006xfi,mao_software_2011}.
Fortunately, with the availability of efficient hardware isolation primitives,
hardware-based vulnerability-isolating systems can guarantee the confidentiality of sensitive kernel data
without incurring substantial performance overhead~\cite{Narayanan2019,narayanan2020lightweight,mckee2022preventing}.

\noindent{\bf{\emph{Integrity.}}}
An adversary can abuse write primitives of a compromised kernel component
to corrupt critical kernel state (e.g., a page table) or tamper with run-time checks.
Therefore, vulnerability-isolating systems provide additional integrity guarantees,
on top of data and control-flow integrity discussed in~\autoref{sec:categorisation:exclave:faultisolation},
to protect security-critical code and data.
For example,
both BGI~\cite{castro2009fast} (a fault-isolating system)
and XFI~\cite{erlingsson2006xfi} (a vulnerability-isolating system)
instrument run-time checks
in untrustworthy kernel components
to enforce isolation.
However,
XFI additionally verifies the integrity
of its checks
at load time
to prevent
an adversary with an arbitrary write primitive
from modifying these checks.

\subsection{Safebox Architecture}
\label{sec:categorisation:enclave}

Attackers can exploit kernel vulnerabilities to bypass security mechanisms
(\eg ASLR~\cite{cve-2019-11190,cve-2015-1593,cve-2014-9585} and SELinux~\cite{cve-2016-10044}).
The safebox architecture isolates security-critical kernel components,
such as security-related kernel state %
and in-kernel policy enforcement mechanisms,
from the rest of the monolithic system to prevent security bypass.
To do so,
it restricts resource accesses and control transfers
\emph{into} an isolated compartment.

\noindgras{Resource Access Control.}
The safebox architecture compartmentalizes security-related kernel resources
and restricts their~access
to only the trusted code within the compartment.
Since the rest of the kernel has no access to the isolated kernel data,
its integrity is preserved even in the presence of a compromised kernel.

\noindgras{Control Transfer Restriction.}
A safebox system
allows protected kernel operations
to execute only within the safebox.
The rest of the kernel transfers control into the safebox whenever a protected kernel operation is requested.

\noindgras{Use Cases and Security Properties.}
Kernel security mechanisms can be protected under two threat models:
(1) the kernel is benign but contains \emph{exploitable} vulnerabilities;
(2) the kernel is compromised and potentially \emph{malicious}.
In the \emph{exploitable} threat model, an unprivileged attacker can exploit kernel vulnerabilities to achieve privilege escalation.
The \emph{malicious} threat model assumes that a kernel has fallen under the control of an attacker, \ie kernel data and kernel code can be modified to exhibit malicious behavior.
Safebox systems defend against the \emph{exploitable} kernel through \emph{isolation of security-critical state}, and protect against a \emph{malicious} kernel by providing an isolated execution environment for \emph{secure in-kernel policy enforcement}.
We discuss the differences in their \emph{integrity} and \emph{confidentiality} guarantees below.

\subsubsection{Security State Isolation}
\label{sec:overview:safe:state}

Attackers can exploit kernel vulnerabilities
to obtain arbitrary read/write primitives~\cite{cve-2013-6282},
which enable them to corrupt security-critical state to bypass security mechanisms.
For example, KCoFI leverages hardware trap vector tables
and page mapping information to enforce control-flow integrity~\cite{criswell2014kcofi}.
However,
attackers can bypass this ad-hoc security mechanism
by corrupting this security-critical state.
Therefore,
to protect these security mechanisms,
safebox systems isolate such state,
enforcing \emph{confidentiality} and/or \emph{integrity}.

\noindent{\bf{\emph{Confidentiality.}}}
The leakage of kernel state
could lead to security bypass,
so confidentiality is important
to safeguard defense mechanisms.
For example,
prior attacks~\cite{kaslr_bypass1,kaslr_bypass2,kaslr_bypass3}
exploit memory safety bugs,
which reveal internal kernel state,
to bypass kernel address space layout randomization (KASLR).
Safebox systems preserves confidentiality by restricting the kernel's read access to data stored within a safebox.
However, similar to sandbox systems, software-based safebox systems~\cite{criswell2014kcofi,song2016enforcing} often do not guarantee confidentiality
due to performance concerns
while hardware-based ones~\cite{proskurin2020xmp,iskios_raid_21} do.

\noindent{\bf{\emph{Integrity.}}}
To prevent tampering of security-critical state,
safebox systems restrict the kernel's write access to the safebox.
For safebox systems that rely on the memory management unit (MMU) for correct configuration of access permissions or page table mappings, a corrupted MMU can violate safebox isolation.
Therefore, MMU integrity must also be preserved in such systems~\cite{criswell2014kcofi,song2016enforcing,proskurin2020xmp,iskios_raid_21}.

\subsubsection{Secure Policy Enforcement}

When a kernel is compromised, security mechanisms are at risk because attackers can arbitrarily modify kernel state and code.
Although the state (data) of a security mechanism can be isolated with security-state-isolating safeboxes, the policy (code) itself can still be corrupted.
Existing safebox systems~\cite{dautenhahn2015nested,azab_skee_2016} leverage
compartmentalization to provide a safe execution environment
for policy enforcement,
in which the TCB no longer includes the entire kernel.
In other words, the \emph{integrity} of the isolated security mechanism is not affected even in the presence of a compromised kernel.

\noindent{\bf{\emph{Integrity.}}}
Similar to security state isolation, the kernel is restricted from writing to the safebox, and MMU isolation is required if it is involved in compartmentalization.
However, the MMU isolation strategy differs. %
In security state isolation (\autoref{sec:overview:safe:state}), the MMU management code, though sometimes instrumented with run-time checks~\cite{criswell2014kcofi}, executes outside the safebox so the kernel retains control over the kernel memory.
This MMU isolation strategy works because the kernel is assumed benign.
However, secure-policy-enforcement systems assume a malicious kernel, so leaving the MMU in the hands of a malicious kernel could break MMU isolation and potentially the safebox itself.
Therefore, these systems deprive the kernel of critical MMU functionalities so it cannot modify the MMU without switching to the safebox.
This allows the safebox to gain exclusive control over the entire kernel memory, preventing the malicious kernel from bypassing security mechanisms by e.g., changing access permissions or injecting malicious code.

\subsection{Summary}

Our categorization
of sandbox and safebox architectures
is broad,
abstracting away nuanced differences
among a wide variety of threat models
adopted in the literature
to emphasize their high-level commonalities.
However,
\emph{the devil is in the details}:
With almost
no published work
making the exact same threat assumptions,
a direct comparison between two systems
is unequivocally difficult --
an important but unfortunate
aspect of kernel compartmentalization work
that we will revisit in~\autoref{sec:evaluation}.
It is therefore crucial
to examine
\emph{explicit} and \emph{implicit} threat assumptions
made by authors
when identifying a suitable mechanism
for a given use case.

\section{Compartment Boundaries}
\label{sec:boundaries}
Kernel compartmentalization involves 1) creating compartments
and 2) enforcing isolation.
In this section,
we describe the process of 
compartment creation,
which requires \emph{kernel decomposition}
(to separate the code and data of a kernel component to be compartmentalized)
and \emph{glue code generation}
(to marshal and synchronize data between an isolated component 
and the rest of the kernel).
We discuss the challenges 
in both steps and
solutions proposed by existing kernel compartmentalization systems 
to address them. 

\subsection{Kernel Decomposition}
\label{sec:boundaries:decomposition}

Prior work on software compartmentalization (\autoref{sec:background:software}) 
splits a user-space application into a privileged and an unprivileged component.
This process,
known as \emph{privilege separation},
can be done either manually~\cite{mccune2010trustvisor,kilpatrick2003privman,ta2006splitting} or automatically~\cite{brumley2004privtrans,lind2017glamdring,bittau2008wedge,liu2019program,Liu2015}.
Manually compartmentalizing any non-trivial legacy application
is impractical,
due to \eg extensive uses of function pointers 
(such as in OpenSSL).
Work such as Privtrans~\cite{brumley2004privtrans}, Wedge~\cite{bittau2008wedge}, and SeCage~\cite{liu2019program} instead
leverages program analysis techniques to automatically partition user-space applications.
However, the techniques they use cannot be applied directly to kernel code for the following reasons:

\noindgras{Kernel Size:}
Current program analysis techniques,
regardless of whether they can be automated or not,
do not scale well
to the large code base
of a monolithic kernel.
This is because program analysis rapidly becomes prohibitively expensive
as the number of program states
exponentially increases~\cite{hind2001pointer,tok2006efficient,zuo2021systemizing}.
Although scalability can be improved at the expense of soundness, which is a reasonable trade-off for bug-finding tools~\cite{machiry2017dr},
an unsound analysis in the context of partitioning could lead to correctness issues 
if some critical data are left unprotected. 
Thus, this trade-off must be considered carefully
in kernel space. 

\noindgras{Complex Kernel Data Dependencies:}
Kernel compartmentalization must enforce 
restrictions on all resources required by an isolated kernel component,
including resources shared between the compartment and the rest of the kernel.
For correctness and performance, decomposition frameworks must 
infer the component's
data dependencies soundly and precisely.
However, pointers in data structures make precisely identifying decomposition boundaries difficult.
To understand pointer aliasing,
a common form of data dependency in C/C++,
these frameworks must perform complex global pointer analysis,
which does not scale to large programs~\cite{liu2017ptrsplit}.
As such, 
many software compartmentalization systems~\cite{brumley2004privtrans,bittau2008wedge,liu2019program} 
lack support for dependency analysis involving pointers;
others~\cite{ganapathy_microdrivers_2008,renzelmanndecaf} often over-estimate shared resources
to ensure soundness,
thus impacting performance
(\eg as a result of unnecessary synchronization)~\cite{huang2022ksplit}.

\noindgras{Multi-threading:}
Most user-space decomposition mechanisms~\cite{brumley2004privtrans,gudka2015clean,lind2017glamdring, liu2019program} do not support multi-threading, 
but it is a key feature in commodity kernels.
Moreover, 
multi-threading introduces concurrency issues
that must be addressed with synchronization mechanisms during glue code generation~(\autoref{sec:boundaries:codegen}).

\vspace{2mm}
\noindent Depending on the architecture (\autoref{sec:categorisation}),
decomposition focuses on different parts of the kernel.
In the \emph{sandbox} architecture,
a kernel compartmentalization system 
typically performs
\emph{device driver decomposition},
while \emph{security state decomposition} 
and \emph{policy enforcement decomposition}
are usually the focus of a \emph{safebox} system.

\subsubsection{Device Driver Decomposition}
Device driver decomposition techniques can be divided into \emph{manual} and
\emph{automated} approaches. The majority of published
research~\cite{witchel2005mondrix,swift2003improving,safedrive06,castro2009fast,mao_software_2011,Narayanan2019,narayanan2020lightweight,mckee2022preventing}
manually decomposes device drivers.  For example, LXFI~\cite{mao_software_2011}
supports fine-grained privilege separation by allowing users to manually
annotate kernel interfaces and specify security policies for each instance of a
shared kernel module (\eg a block device).  If a module instance is compromised,
the attacker can only exploit the privileges associated to the instance (\eg
write only to the affected block device).  
While LXFI trusts any annotations provided by the user, manual decomposition is
unsustainable and prone to human error,
especially as the size and complexity of driver code increases.

To automate decomposition,
Microdriver~\cite{ganapathy_microdrivers_2008} and Decaf~\cite{renzelmanndecaf} use static analysis
to identify code
that can be delegated to user space,
effectively decomposing a device driver 
into a kernel-level driver (containing high-priority code such as interrupt handling) 
and a user-level driver (containing infrequently invoked, low-priority code),
which allows the user-level driver 
to be written in safer languages than C
(\eg Java~\cite{renzelmanndecaf} in Decaf).
This approach resembles 
the typical mechanism/policy split in microkernel architectures 
(\autoref{sec:background:alternatives}).
However,
only a subset of driver functionality is moved to user space, 
so faults remaining in the kernel part of the decomposed driver can still pose a threat to the system's reliability.
To scale the decomposition framework to the entire device driver, 
program analysis approaches need to overcome the aforementioned challenges (e.g., alias analysis).
For example,
KSplit~\cite{huang2022ksplit} addresses
the scalability issue faced by
many user-space privilege separation techniques~\cite{lattner2007making,sui2016svf}
via the parameter tree approach~\cite{liu2017ptrsplit},
which is a modular way of computing dependence graphs
from programs containing complex uses of pointers.

\subsubsection{Security State Decomposition}
To isolate security-critical kernel state (\eg page tables and process credentials),
some work~\cite{criswell2014kcofi,proskurin2020xmp,iskios_raid_21}
\emph{manually} identifies the critical data of a security mechanism
and stores them 
in a protected memory region.
For example, KCoFI~\cite{criswell2014kcofi} isolates a region of memory reserved for storing security-sensitive data 
like page mapping information.

While manual decomposition is suitable
for well-scoped kernel data, 
automated static analysis is necessary
when the data to be isolated is non-trivial to infer 
(\eg when the sensitive data is mixed with other kernel data). 
For example, 
KENALI~\cite{song2016enforcing} protects access control checks 
scattered throughout the kernel 
by isolating all data that is relevant to the checks.
It leverages the implicit semantics of security checks,
which specify that
every check must return a security-related error code when it fails.
The semantics help KENALI reduce its search space to only the functions 
that could return a ``permission denied'' error code.
A sound dependency analysis
to ensure that no sensitive data is left unprotected
then becomes feasible,
as long as the kernel code strictly adheres
to the semantics.
	 
\subsubsection{Policy Enforcement Decomposition}
Unlike device drivers and security-critical state, 
a host's policy enforcement mechanism is often \emph{directly} loaded into a secure compartment
during secure boot,
\emph{without} the need for any decomposition analysis.
For instance, PerspicuOS~\cite{dautenhahn2015nested} and SKEE~\cite{azab_skee_2016} provide a secure %
environment 
where the policy enforcement mechanism 
can safely execute 
even %
in a %
malicious kernel.
Since a secure compartment already 
isolates the %
policy enforcement mechanism from the kernel,
the mechanism itself 
does not need to be decomposed
to protect its critical~data.

\subsection{Glue Code Generation}
\label{sec:boundaries:codegen}
The main purpose of ``glue code'' is 
to \emph{marshal} and \emph{synchronize} data 
across compartment boundaries.
Marshaling is necessary,
because some isolation techniques (\autoref{sec:mechanisms})
place a kernel compartment into a separate virtual address space, 
so a cross-compartment procedure call
is needed to execute a function in another address space.
Synchronization %
is necessary,
because all Unix kernels are reentrant~\cite{bovet2005understanding}, \ie
multiple kernel processes can safely execute concurrently without introducing
any consistency issues among kernel resources.
Shared resources and synchronization primitives (\eg spinlocks and mutexes) 
must be
updated upon every function invocation 
that crosses the compartment boundary.
We discuss glue code generation 
for each type of decomposition we presented in~\autoref{sec:boundaries:decomposition}.

\subsubsection{Device Driver Decomposition}
Most software-based kernel compartmentalization systems~\cite{safedrive06,castro2009fast,erlingsson2006xfi,mao_software_2011}
isolate device drivers in a \emph{logically separate} portion of the kernel address space,
so they remain in the same virtual address space as the kernel.
These systems typically enforce isolation via run-time checks on read, write, and jump instructions,
restricting access to arbitrary kernel memory while sharing with the rest of the kernel a single copy of kernel data.
Unless these checks require additional information
from the kernel that must be 
updated upon cross-compartment transitions
(\eg in BGI~\cite{castro2009fast}, an access control list is used
for permission checking; see also~\autoref{sec:mechanism:intrakernel-domain}),
data marshaling and synchronization %
is generally not necessary.
On the other hand, Nooks~\cite{swift2003improving} requires glue code, 
but only for data synchronization,
because it keeps a shadow copy of shared kernel resources
for isolated device drivers to directly modify;
marshaling is not needed, 
because kernel pages are mapped into a compartment's page table for read access.

Glue code can be generated \emph{manually}~\cite{witchel2005mondrix,swift2003improving,nikolaev_virtuos_2013,mckee2022preventing} or \emph{automatically}~\cite{ganapathy_microdrivers_2008,renzelmanndecaf,Narayanan2019,narayanan2020lightweight}.
Nooks~\cite{swift2003improving} does so manually;
developers implement synchronization routines 
to copy shared resources between an isolated device driver and the kernel.
Microdriver~\cite{ganapathy_microdrivers_2008} and Decaf~\cite{renzelmanndecaf} leverage field access analysis, a static
analysis technique that identifies the fields of a complex data
structure crossing the compartment boundary,
to automatically identify data fields to be marshaled and synchronized.
However, as discussed in~\autoref{sec:boundaries:decomposition}, static analysis
is complicated by the proliferation of pointers in the kernel.
For example, C/C++ pointers do not carry bounds information, 
making data marshaling difficult~\cite{liu2017ptrsplit}. %
To ensure completeness,
Microdriver and Decaf still rely on user annotations
to understand the semantics of
pointers in kernel data structures
(\eg to determine the size of an array). \emph{Automated marshaling} of
pointer data has subsequently been made possible by PtrSplit~\cite{liu2017ptrsplit}.
PtrSplit instruments the compartmentalized program to track the bounds of pointers, 
so that all pointers that cross the compartment boundary have the bounds information necessary for automatically marshaling pointer data.

LXDs~\cite{Narayanan2019} and LVD~\cite{narayanan2020lightweight} 
do not use static analysis
but instead define an Interface Definition Language (IDL), 
a special-purpose language
for describing interfaces between software components, 
to automatically generate glue code. %
However, 
they still require 
developers to manually define
the IDL specification for device drivers.
To reduce the effort involved,
KSplit~\cite{huang2022ksplit} 
uses static analysis
to generate IDL specifications for marshaling and synchronizing data across compartment boundaries;
user annotations are needed
only when
the analysis
cannot resolve ambiguities (\eg unions or void pointers).
Once a working IDL specification is defined
(either manually in LXDs and LVD,
or semi-automatically in KSplit),
the IDL compiler can automatically generate the glue code.

\subsubsection{Security State and Policy Enforcement Decomposition}
Sharing protected data 
between a compartment
and the kernel is disallowed \emph{by design} in mechanisms that isolate security state or support secure policy enforcement,
because the goal of a safebox compartmentalization system~\cite{criswell2014kcofi,song2016enforcing,proskurin2020xmp,dautenhahn2015nested,azab_skee_2016,iskios_raid_21}
is to ensure that the protected memory can \emph{only} be accessed by the trusted code in the isolated compartment.
Therefore,
such a system does not require
any glue code for data marshaling or synchronization,
irrespective of the isolation technique 
used for compartmentalization.

\subsection{Summary}

A monolithic kernel is typically built from %
a large and complex code base.
Dependencies between sub-components are difficult to identify,
frequently hidden behind programmatic abstractions.
Relying completely on manual decomposition effort 
is impractical,
especially given the sheer amount of code to analyze 
and the large number of modifications to the code base 
required by compartmentalization
(\eg inserting synchronization primitives).
Meanwhile, 
fully automated approaches also fall short of expectations. %
\emph{Finding a middle ground
with acceptable trade-offs seems inevitable:}
We must continue to improve
automated program analysis techniques
to identify compartment boundaries
and insert necessary code for across-boundary interactions,
and involve humans in the loop
only when code complexity
exceeds the capacity of current analysis techniques.
While advances in automated techniques
continue to reduce the amount of
manual intervention required,
it is unclear from present work %
how often humans are involved in practice.
We further discuss this issue in~\autoref{sec:evaluation}.

\section{Compartment Isolation Mechanisms}
\label{sec:mechanisms}
A kernel compartmentalization system
must enforce an isolation policy
on a decomposed kernel component
to control resource accesses
and restrict control transfers.
However, unlike user-space isolation,
where policy enforcement typically runs
at a higher privilege level
than the isolated user-space code~\cite{Liu2015,karande2017sgx,paccagnella2020custos,narayan2021swivel},
kernel isolation faces
the challenge that
the isolated component executes in the same privileged kernel mode
as the rest of the kernel.
Since kernel privilege grants an attacker
access to critical kernel state %
and code in memory,
enforcing kernel isolation
requires some form of privilege separation
from the decomposed kernel component
to prevent security bypass.
Existing %
work leverages the \emph{user/kernel}~(\autoref{sec:mechanism:userkernel})
or \emph{kernel/hypervisor}~(\autoref{sec:mechanism:hypervisor})
privilege separation
to enforce isolation at a higher privilege level
than the decomposed kernel component.
Isolation can also be enforced at the same privilege level
(\ie \emph{intra-kernel} isolation)
using either
\emph{address-based}~(\autoref{sec:mechanism:intrakernel-address}) or
\emph{domain-based}~(\autoref{sec:mechanism:intrakernel-domain})
techniques.
We categorize isolation mechanisms
based on these privilege separation models
and discuss how they enforce
sandbox- and safebox-based kernel compartmentalization.

\subsection{User/Kernel}
\label{sec:mechanism:userkernel}

Commodity OSs leverage processor modes (\eg user and kernel mode)
to enforce privilege separation.
All user code
runs in a separate virtual address space
and must rely on the kernel via system calls
to execute privileged instructions.

\begin{figure*}[t]
	\begin{subfigure}[b]{0.48\textwidth}
		\centering
		\includegraphics[width=\textwidth]{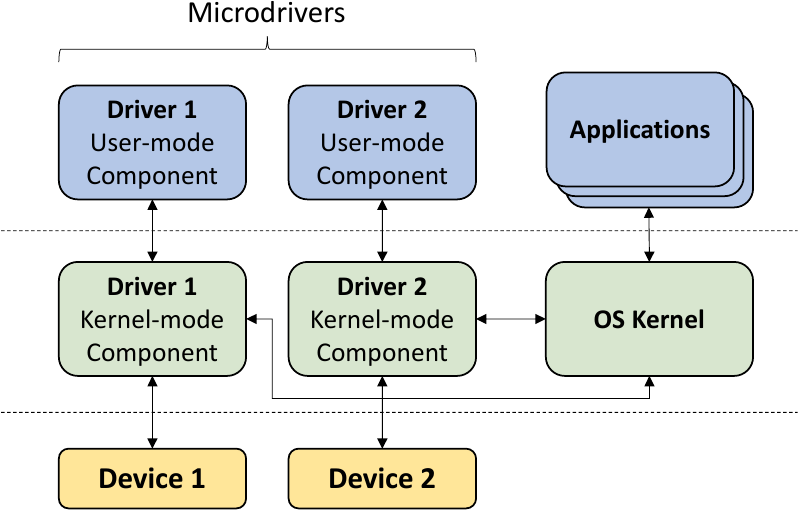}
		\caption{Microdriver Architecture~\cite{ganapathy_microdrivers_2008}}
		\label{fig:microdriver}
	\end{subfigure}
	\hfill
	\begin{subfigure}[b]{0.48\textwidth}
		\centering
		\includegraphics[width=\textwidth]{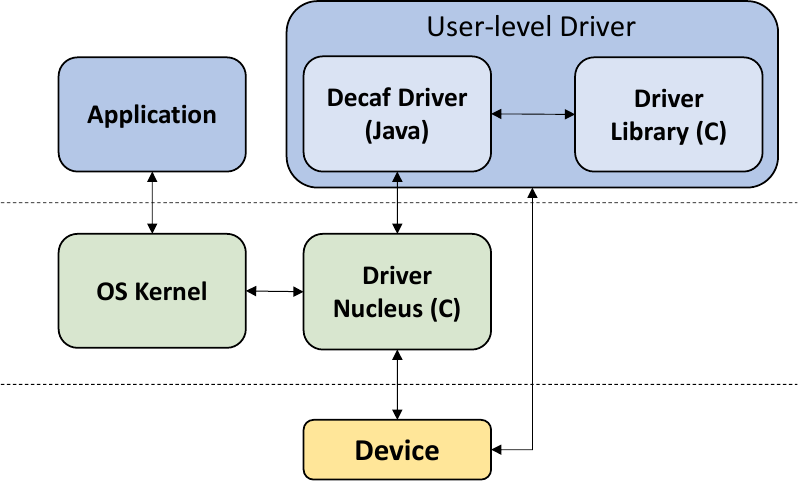}
		\caption{Decaf Architecture~\cite{renzelmanndecaf}}
		\label{fig:decaf}
	\end{subfigure}
	\caption{Different user/kernel compartmentalization approaches. In Microdriver, the user-mode driver is written in the same C language as the kernel-mode driver; in Decaf, the user-mode driver can be written in languages other than C.}
	\label{fig:user-kernel}
\end{figure*}

\noindgras{Sandbox-based Isolation.}
The hardware-enforced user/kernel boundary
can be used to sandbox untrusted kernel code
(\eg device drivers) in various ways. %
At one extreme, early work %
effectively converts a monolithic kernel
into a microkernel architecture (\autoref{fig:kernel_architectures}).
For example,
Sawmill~\cite{Smith1985} %
partitions
the monolithic kernel %
into a core kernel and a number of %
services,
and moves these services to user space.
However, switching between the user- and kernel-mode code is costly,
especially for high-throughput devices such as network cards.
To address this issue,
Microdriver~\cite{ganapathy_microdrivers_2008} sandboxes only a subset of device driver functionality
(\autoref{fig:microdriver}),
relegating rarely used code (\eg device startup and shutdown)
to user space
while leaving performance-critical code (\eg data transmission)
in the kernel.
The performance gain comes at the cost of security,
since the kernel-mode code,
which might contain faults to corrupt the kernel,
continues to execute in the same address space
as the rest of the kernel.
However, unused or less-frequently-used code -- which is isolated by Microdriver --
is the subset of the code base that is known
to contain more bugs~\cite{kurmus2011attack,azad2019less,ghaffarinia2019binary}
and thus more likely to be leveraged as an attack vector.
Renzelmann~\etal~\cite{renzelmanndecaf} take this approach one step further,
implementing user-space code in a safer language
(Java in their prototype, see~\autoref{fig:decaf})
that is less prone to memory vulnerabilities common in C.

\noindgras{Safebox-based Isolation.}
Many user-space
compartmentalization systems~\cite{karande2017sgx,paccagnella2020custos}
leverage Trusted Execution Environments (TEEs)
to protect applications
from a malicious kernel.
For example,
Intel's Software Guard Extensions (SGX)~\cite{costan2016intel},
a widely used TEE,
uses hardware-based cryptography
to protect the confidentiality and integrity
of user-mode code
from the rest of the system.
However,
kernel functionality
cannot execute within an SGX enclave;
moreover,
as with sandbox-based isolation,
any communication between an enclave and
trusted kernel functionality
introduces large overhead.
Therefore,
we expect that
user-space TEEs such as SGX
will remain of limited practical use for
compartmentalizing kernel functionality.
However, as we discuss in \autoref{sec:mechanism:hypervisor}, ARM's TrustZone
technology~\cite{tzone} has been used for kernel compartmentalization.

\subsection{Kernel/Hypervisor}
\label{sec:mechanism:hypervisor}

Hypervisors run at a higher privilege level than the kernel,
managing %
virtual machines (guest OSs)
that share %
physical computing resources. %
Kernel compartmentalization %
isolates a kernel component within a VM,
which executes in a separate virtual address space,
while enforcing security policies from
the higher-privileged hypervisor.
Unlike user/kernel isolation~(\autoref{sec:mechanism:userkernel}),
compartments in kernel/hypervisor isolation
execute in privileged kernel mode
and therefore have direct access to privileged instructions.
However,
since attackers can abuse privileged instructions to break the isolation,
a compartment is subject to
some restrictions on its use of those instructions,
whereby protected sensitive instructions
(\eg programming of Interrupt Controllers~\cite{narayanan2020lightweight})
require a hypercall into the hypervisor.
We discuss techniques~\cite{Narayanan2019,narayanan2020lightweight}
that mask and thus reduce
the cost of transitioning across the kernel/hypervisor boundary in~\autoref{sec:mechanism:optimizations}.

\begin{figure*}
	\centering
	\begin{subfigure}[b]{0.24\textwidth}
		\centering
		\includegraphics[width=0.9\textwidth]{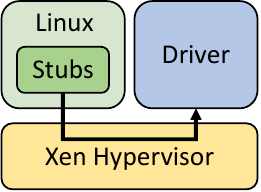}
		\caption{Xen split-driver architecture}
		\label{fig:xen}
	\end{subfigure}
	\hfill
	\begin{subfigure}[b]{0.24\textwidth}
		\centering
		\includegraphics[width=0.9\textwidth]{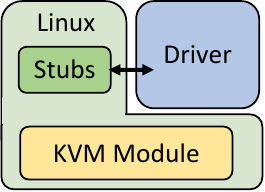}
		\caption{LXDs architecture~\cite{Narayanan2019}}
		\label{fig:lxds}
	\end{subfigure}
	\hfill
	\begin{subfigure}[b]{0.24\textwidth}
		\centering
		\includegraphics[width=0.9\textwidth]{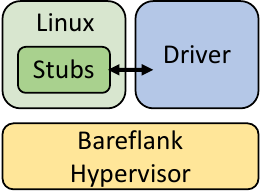}
		\caption{LVD architecture~\cite{narayanan2020lightweight}}
		\label{fig:lvd}
	\end{subfigure}
	\hfill
	\begin{subfigure}[b]{0.24\textwidth}
		\centering
		\includegraphics[width=0.9\textwidth]{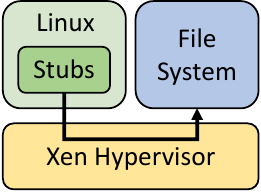}
		\caption{VirtuOS architecture~\cite{nikolaev_virtuos_2013}}
		\label{fig:virtuos}
	\end{subfigure}
	\caption{Different hypervisor-based compartmentalization approaches.}
	\label{fig:hypervisors}
\end{figure*}

\noindgras{Sandbox-based Isolation.}
The Xen hypervisor's
\emph{split-driver} model  (\autoref{fig:xen})
is an early example of
using virtualization
for %
kernel compartmentalization~\cite{pratt2005xen}.
In this model,
a front-end driver
running in a \texttt{DomU} VM
forwards all driver requests
to a back-end driver
that can interact with the physical device.
By default,
the back-end driver runs in the \texttt{Dom0} VM,
a privileged VM started by the hypervisor
to manage \texttt{DomU} VMs.
However,
a faulty back-end driver in \texttt{Dom0}
can affect the hypervisor;
moreover,
it cannot easily \emph{recover} from an error,
since the \texttt{Dom0} VM cannot be restarted.
To improve \emph{reliability} and \emph{recoverability},
the split-driver model allows
the back-end driver
to run in a separate \texttt{DomU} VM,
called the \emph{driver domain}~\cite{xenDriverDomain},
so that a faulty driver
can be killed or restarted independent of other VMs
(including the one running the trusted kernel).

LXDs~\cite{Narayanan2019} (\autoref{fig:lxds}) use
the KVM hypervisor~\cite{linuxKVM},
instead of Xen,
to compartmentalize device drivers.
Like in Xen's split-driver model,
LXDs run an untrusted device driver
in a separate VM.
However,
rather than running in a separate VM,
LXDs' trusted kernel
acts as the hypervisor
to manage VMs
running untrusted compartments.

Both the split-driver model and LXDs
incur costs when transitioning between the kernel and hypervisor privilege
levels,
which is more expensive
than transitioning between user and kernel levels
through system calls~\cite{nelson2017hyperkernel}.
To reduce these costs,
LVD~\cite{narayanan2020lightweight} (\autoref{fig:lvd})
uses VMFUNC~\cite{intel_intel_2016}
to enable direct communications
between VMs.
Like in the split-driver model,
the trusted kernel
and untrusted device drivers
run in separate VMs,
but communications between them
do not require going through the hypervisor
(see also~\autoref{sec:mechanism:optimizations}).

Finally, we note that some work has explored ways
to compartmentalize beyond device drivers.
For example,
VirtuOS~\cite{nikolaev_virtuos_2013} (\autoref{fig:virtuos})
runs a complete kernel subsystem
(\eg a network stack or a file system)
in a separate VM.

\noindgras{Safebox-based Isolation.}
The hypervisor can enforce fine-grained memory permissions on VMs
running trusted compartments
by managing multiple \emph{views} of memory.
Each view defines
a set of access permissions for a VM
to restrict code and data accessible
from outside the VM.
xMP~\cite{proskurin2020xmp} uses this approach
to ensure that only trusted compartments
can access security-critical state.
Conceptually,
this approach shares some similarities
with the ones we see in~\autoref{sec:mechanism:intrakernel-address}.

Similar techniques have been applied using ARM's TrustZone~\cite{tzone,pinto2019demystifying}.
For example, Azab \etal~\cite{azab2014hypervision} inserts hooks in the kernel
to transfer control %
to a kernel running in the TrustZone.
The safebox is fully protected from the kernel being monitored and has full access to its memory.
Others~\cite{lentz2018secloak} go a step further
and propose extensions to provide secure access to devices through the safebox,
which can be used to implement I/O reference monitors~\cite{khan2021m2mon,wang2022rt}.

\noindgras{Optimizing Domain Switches.}
As seen in \autoref{sec:mechanism:userkernel} and \autoref{sec:mechanism:hypervisor},
moving kernel code to user space
or a different hypervisor domain
is an attractive way
to compartmentalize a monolithic kernel,
as doing so leverages pre-existing and well-known
hardware-enforced privilege separation.
However,
calling kernel functionality
outside of a compartment
requires an expensive domain switch.
We can minimize this overhead
in three ways~\cite{Narayanan2019,narayanan2020lightweight}:
1) carefully identifying compartment boundaries
to reduce the number of calls (see~\autoref{sec:boundaries}); %
2) using hardware support (see~\autoref{sec:mechanism:optimizations});
and 3) adopting an asynchronous model
and a more efficient IPC mechanism  (see~\autoref{sec:mechanism:optimizations}).

\subsection{Address-based Intra-kernel}
\label{sec:mechanism:intrakernel-address}

\begin{figure}[t]
	\includegraphics[width=\columnwidth]{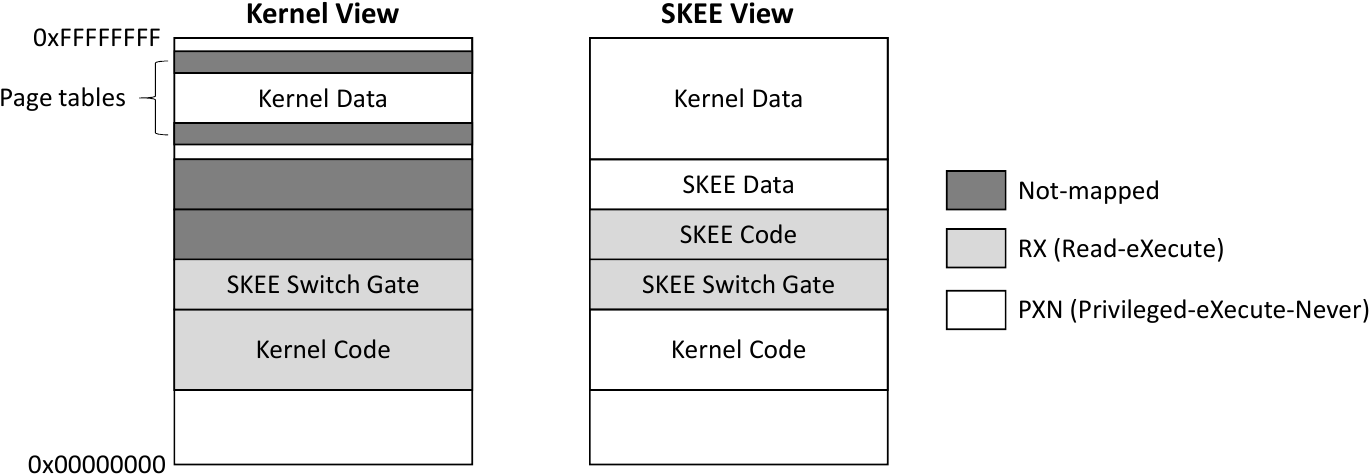}
	\caption{
	An example of address space separation on ARMv7 in SKEE~\cite{azab_skee_2016}
	}
	\label{fig:mmu}
\end{figure}

\emph{Intra-kernel} isolation
means that
an isolated component
runs at the same privilege level
as the rest of the kernel.
As a result,
the compartment
has supervisor privilege
to execute privileged instructions
or access privileged registers.
We classify intra-kernel
approaches
into two types: \emph{address-based} (this section) and
\emph{domain-based} (see~\autoref{sec:mechanism:intrakernel-domain}).

Address-based %
compartmentalization
uses the Memory Management Unit (MMU)
to run a compartment in a separate address space
for isolation.
Since address translation is not shared
between compartments,
an isolated compartment
cannot access data from other compartments.
However, modification to MMU control registers
could allow an untrusted
compartment
to access the data of a trusted compartment.
Therefore,
depending on the threat model,
additional security measures
are required.

\noindgras{Sandbox-based Isolation.}

Nooks~\cite{swift2003improving} leverages the MMU to isolate kernel extensions.
Each compartment has its own set of page tables with appropriate page permissions.
Specifically,
isolated kernel extensions
have read-only kernel access
and read-write access to their own domains
(while the core kernel has read-write access to the entire address space).
Configuring the page table for a domain to remove write access
to the rest of the kernel address space thus suffices to prevent
inadvertent memory corruption.
Note that Nooks is a fault-resistant system
designed to prevent
a buggy extension
from inadvertently corrupting important kernel data structures;
a malicious extension can bypass
its isolation
by reloading the hardware page-table base register.

Since compartments are not permitted to modify kernel data structures directly, Nooks supports cross-compartment interaction
through interposition and object tracking mechanisms that
copy kernel data to compartments and copy them back atomically
after changes have been applied, sanitizing the changes where
possible.
If a check (e.g., parameter validation) fails, or a hardware fault occurs
during domain execution, Nook’s object tracking and recovery
mechanisms release kernel resources held by the domain and
reset the system to a safe state.

\noindgras{Safebox-based Isolation.}
Safebox systems such as PerspicuOS~\cite{dautenhahn2015nested} and SKEE~\cite{azab_skee_2016}
use the MMU to enforce intra-kernel
isolation on commodity hardware
even when the kernel is malicious.
In these systems,
a \emph{trusted} compartment
is responsible for mediating all page table modifications.
Its page table pages
are %
either read-only by the rest of the (potentially malicious)~kernel
or completely inaccessible from outside the compartment.

For example, as illustrated in~\autoref{fig:mmu},
SKEE instruments the kernel's memory translation tables
to 1) unmap entries related to SKEE's isolated environment;
and 2) restrict the kernel's write access to the kernel code
and SKEE's switch gate,
which mediates \emph{all} jumps from the kernel into the isolated environment.
As such,
it safeguards the kernel's only entry point to the isolated environment.

However,
as with Nooks,
a malicious compartment with access to the MMU control registers
can change its %
address translations
to map the trusted compartment's pages.
To prevent this, %
both PerspicuOS and SKEE
scan the kernel code
to remove any instruction
that might be used to modify MMU control registers.
The registers to be protected
are architecture specific.
For example,
PerspicuOS protects %
CR0's write-protection~(WP) bit
on x86/x86-64 and several other control registers,
while SKEE protects ARM's translation table base registers (TTBR0/1) and
the translation table base control register (TTBCR).
We note that a similar approach could be applied to sandbox systems
(like Nooks)
to %
provide stronger isolation
between compartment boundaries.

\subsection{Domain-based Intra-kernel}
\label{sec:mechanism:intrakernel-domain}

Relying on privilege levels to enforce compartment boundaries
comes at the cost of expensive privilege-level transitions.
To avoid this overhead,
researchers have explored ways to enforce restrictions on data accesses
and control transfers within a single address space
at the same privilege level.
Domain-based intra-kernel compartmentalization
divides an address space
into multiple \emph{domains}
that share the same address translation
with each other and the rest of the kernel.
To create a domain,
existing work
proposes both
software- and hardware-based mechanisms
to control access to specific regions of
the virtual memory.
The \emph{granularity} of a domain varies
depending on the mechanism.

\begin{figure}[t]
	\includegraphics[width=\columnwidth]{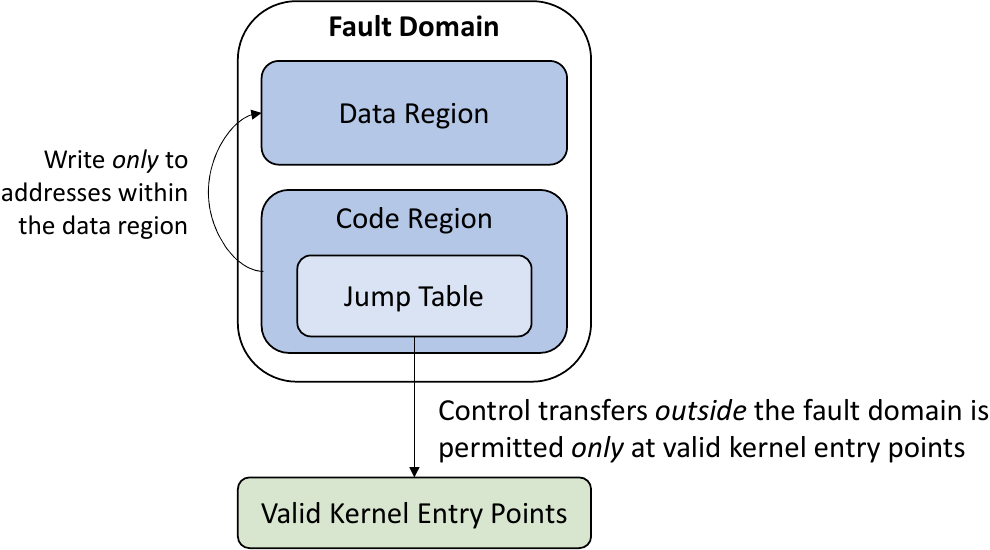}
	\caption{Software fault isolation~\cite{sfi93}}
	\label{fig:sfi}
\end{figure}

\subsubsection{Software-based Isolation}
\label{sec:mechanism:intrakerneldomain:sw}
Software-based %
techniques typically %
instrument untrusted code
with run-time checks
to ensure that
data accesses
and control flow transfers are
within specified bounds as illustrated in \autoref{fig:sfi}.

\noindgras{Sandbox-based Isolation.}
Early work %
built on ideas from \emph{software fault isolation}
(SFI)~\cite{sfi93} to enforce isolation. %
In comparison to address-based
techniques, SFI's run-time checks introduce additional overhead for
intra-compartment execution.
Conversely, SFI avoids costly
page table updates
and data synchronization on compartment transitions~\cite{sfi93}.
Moreover,
SFI enables fine-grained isolation
at the object or even byte level
with minimal or no changes to the existing code and data layout.

For example,
the SafeDrive compiler~\cite{safedrive06} inserts
run-time checks on pointer variables
in untrusted extensions
to enforce type and
memory safety within the extension,
which prevents
spatial bounds violations
in kernel API calls
and direct accesses
to shared kernel objects.
BGI~\cite{castro2009fast} %
creates an Access Control List (ACL) for each byte of memory
to specify access permissions for untrusted extensions. %
ACLs are stored in a protected area in the main memory
to prevent any modification from within an extension.
The BGI compiler instruments
write instructions with run-time checks to restrict
an extension's access to kernel memory. %
Since each run-time check costs additional CPU cycles, to reduce
performance overhead, BGI does not instrument
read instructions;
as a result,
BGI can only guarantee \emph{integrity} but not \emph{confidentiality}.
Trading off confidentiality for performance
is in fact common among SFI-based systems~\cite{erlingsson2006xfi,mao_software_2011}.

\noindgras{Safebox-based Isolation.}
Using software-based isolation techniques
to protect security-critical state
in an untrusted kernel is challenging.
The reason is that the kernel has the privileges to perform
a range of low-level operations that can
violate language-level safety assumptions;
these operations can be abused to corrupt \emph{arbitrary} kernel memory.
Criswell \etal~\cite{criswell_secure_2007} identified
MMU configurations, DMA, and page swapping
as the three kernel functionalities
subject to such manipulation.
They also identified
four types of kernel memory that must be protected
to prevent security bypass:
1) the processor state when held in registers or memory,
2) stack values for kernel threads,
3) memory-mapped I/O locations,
and 4) code pages in memory.

To address some of these challenges, they proposed
secure virtual architecture (SVA)~\cite{criswell_secure_2007}.
SVA is a compiler framework %
that compiles the kernel
to a virtual instruction set
based on LLVM's intermediate representation
and statically checks that
the intermediate bytecode meets several
safety properties (\eg control-flow integrity, type safety for a subset of
objects).
For safety properties that
cannot be guaranteed statically
(e.g., dynamic bounds and indirect calls),
SVA inserts run-time checks.
In addition,
SVA disallows
explicit assembly code in the kernel
(\eg code that handles low-level hardware interactions
to manipulate the MMU);
instead, it requires the assembly code to be replaced
by special SVA instructions~\cite{criswell2006virtual}, implemented as
high-level, well-defined intrinsic functions, to enforce memory safety guarantees.
For example, the \lstinline[language=C,basicstyle=\ttfamily]{sva.save.integer(void* buffer)} function is used to
save the processor's integer \emph{control state} (\ie control and general-purpose registers)
into the memory pointed to by
\lstinline[language=C,basicstyle=\ttfamily]{buffer} (\eg on a context switch).

KCoFI~\cite{criswell2014kcofi} uses SVA to %
enforce \emph{control-flow integrity (CFI)} on kernel code.
To protect the security-critical state of the CFI mechanism
(\eg page mapping information),
KCoFI stores them in an isolated memory region. %
KCoFI instruments all kernel store instructions
to prevent errant writes %
from corrupting them.
To ensure the integrity
of the kernel code and the address space layout,
KCoFI uses SVA to restrict
how the kernel configures the MMU.
Further, it relies on SVA
to protect against DMA attacks that
manipulate the IOMMU
and several other
low-level, security-sensitive kernel operations and states
as identified by Criswell~\etal~\cite{criswell_secure_2007}.

\subsubsection{Hardware-based Isolation}
\label{sec:mechanism:intrakerneldomain:hw}
Hardware features from different CPU vendors vary,
but existing domain-based work typically leverages them
to provide page-sized or even finer-grained
(\eg byte-level) isolation.
Some hardware features can be used
in both sandbox and safebox
architectures %
to optimize the techniques mentioned
in software-based isolation (\autoref{sec:mechanism:intrakerneldomain:sw}).
Therefore,
we organize the remainder of this section
based on the granularity of the protection,
instead of on the sandbox/safebox architecture.

\mypar{Page-sized Isolation}
\emph{Memory Protection Keys} (MPK)
were introduced in Intel's Skylake processor~\cite{intel_intel_2016}
to separate the virtual address space
into %
16 domains %
using four designated bits in every page table entry.
Each CPU core has a special \texttt{pkru} register
that stores the core's permissions for every domain.
However,
while only the kernel can modify a page's domain
(since it is stored in the page table),
the \texttt{pkru} register can be modified in user mode.
If MPK is used for kernel compartmentalization,
then any user process can invalidate
compartmentalization
by changing the value of the \texttt{pkru} register.
To circumvent this limitation,
IskiOS~\cite{iskios_raid_21}
reserves eight domains
for kernel compartmentalization
and allows user-space applications
to use only the remaining domains.
When a user process attempts to
write to the \texttt{pkru} register,
IskiOS traps to the kernel,
which intercepts the instruction
and checks that only user-space domains
are modified.
However,
user applications running on IskiOS
incur performance overhead
on every write to \texttt{pkru}. %
It is important to note that IskiOS assumes that the kernel might have buffer overflow bugs,
but is not compromised by the attacker.

Recently, Intel announced \emph{Supervisory Keys (PKS)}~\cite{PKSPMEMAdd} which adds a new \texttt{pkrs}
register that is accessible only from the kernel mode.
PKS can be used to implement kernel compartmentalization, just as described by
IskiOS, but without trapping on writes to the \texttt{pkru} register and instead using the \texttt{pkrs}
register.
At the time of this writing,
however, no system has used PKS for kernel compartmentalization.

Similar to MPK, \emph{Memory Domains} (MD)
on ARM-v7 CPUs enables an address space to be split into compartments.
The kernel assigns every page to one of 16 domains
using a 4-bit flag in the Page Directory Entry (PDE).
The Domain Access Control Register (DACR),
similar to the \texttt{pkru} register,
dictates each core's permission for every domain.
However,
unlike the \texttt{pkru} register,
both PDE and DACR can be accessed only in kernel mode.
Therefore, MD-enabled kernel compartmentalization seems plausible,
but prior work~\cite{chen2016shreds} has used
MD to create user-space compartments only.
Similar to IskiOS, such an approach trusts parts
of the kernel to set up these domains.

\mypar{Finer-grained Isolation}
Mondrix~\cite{witchel2005mondrix} proposes custom hardware
extensions to support fine-grained compartmentalization
of the Linux kernel.
Mondrix builds on %
Mondrian~\cite{witchel2002mondrian},
which supports flexible, controlled memory-sharing
between isolated domains in user space.
Mondrian associates
a domain to a permission table,
which describes
word-level access permissions
in a process' address space.
Each core has a register
that points to the permission table
associated with the current domain
and a dedicated lookaside buffer for performance.
Permission tables can only be edited
by a privileged memory supervisor running in domain 0.
	Mondrix adapts Mondrian's memory protection mechanisms to the Linux
	kernel.
	As with Mondrian, only the memory supervisor can access
	all of the memory without the mediation of a permissions table.
	There is one memory supervisor in the kernel, and it is trusted.
	To improve performance of
	domain switches, Mondrix stores context information (e.g., stacks) in user
	memory that is writeable by hardware but read-only to software other than the
	memory supervisor. This reduces the overhead from context switching between
	kernel threads on cross-domain calls. Mondrix also adds hardware to
	protect stacks of different threads resident in the same protection domain. Apart from the need for custom hardware, a disadvantage of
	Mondrix is the need for a separate protection table for each domain, which may
	limit domain scalability. In addition, although it provides finer granularity than that of MMU approaches (\autoref{sec:mechanism:intrakernel-address}),
	Mondrix still only provides limited protection for sub-word granularity
	allocations (\eg array entries or stack frames).

To support byte-level memory protection,
Intel introduced \emph{Memory Protection Extensions} (MPX)~\cite{intel_intel_2016} in 2015.
With MPX, programmers can create and enforce bounds by specifying two 64-bit
address denoting the start and end of a memory range.
Two new instructions,
\texttt{bndcl} and \texttt{bndcu},
perform bounds checking on an address.
For \compzz, the address space is partitioned using MPX bounds.
By defining a single bound and adding a single bound check before
every memory access, we ensure that every access is within the allowed
partition on the address space.
However, MPX's bounds registers (\texttt{bnd0} to \texttt{bnd3})
 can store at most four bounds;
 additional bounds must then be stored in memory,
which affects performance~\cite{memsentrykoning2017no}.
Therefore, MPX-based approaches can only efficiently
support four domains, %
and the performance degrades
with more domains.
Koning~\etal~\cite{memsentrykoning2017no}
show that MPX outperforms SFI in most cases,
but since this work is done only for user-space compartmentalization,
the applicability of their results to
kernel space remains to be confirmed.
Note that MPX was discontinued in 2019
due to performance overhead~\cite{oleksenko2018intel} and
side-channel~vulnerabilities~\cite{mpxsidechannel}.

ARM's Memory Tagging Extensions (MTE)~\cite{arm_mte} and Pointer Authentication
Codes (PAC)~\cite{arm_pac} are new security features on ARM processors. With
respect to kernel compartmentalisation, MTE is similar to both Intel MPX in that it
supports fine-grained memory protection, and to Intel MPK in that it associates a
domain with a virtual memory region. Unlike MPK, its granularity is much smaller
(16 bytes instead of 4KB). In MTE, each protected memory region is tagged
with one of up to 16 \emph{colors}. A straightforward application of MTE to
kernel compartmentalisation is to tag each domain with a unique color. However,
as with MPK, this approach suffers from limited domain scalability.

HAKC~\cite{mckee2022preventing} combines PAC and MTE to overcome the scalability
limitations of MTE for kernel compartmentalisation. PAC introduces new instructions for code and data \emph{pointer authentication}
to protect the integrity of code and/or data pointers. On pointer creation, PAC
computes a cryptographic message authentication code (MAC), referred to as a
\emph{pointer authentication code} (PAC), over the pointer and additional
non-secret context data. The PAC is then stored in unused higher-order bits of
the pointer.
Subsequent attempts to load and/or dereference the pointer can be
instrumented with instructions to check whether the pointer still matches its PAC and
has not been corrupted by an adversary. PAC does not provide full memory
safety and is potentially vulnerable to malicious PAC generation and reuse
attacks~\cite{pacitup}. Nonetheless, precise code pointer integrity~\cite{cpi}
prevents all known control-flow hijacks, and data pointer integrity protects
against many powerful data-oriented attacks (\eg~\cite{dop}).

To combine PAC with MTE, HAKC proposes a two-level design where compartments are subdivided into
up to 16 different \emph{cliques}. Cliques can access data in other cliques in
the same compartment, subject to a \emph{clique access policy}, but are
prevented from accessing other compartments. Control transfers between
compartments are restricted according to a \emph{compartment transition policy},
and can only jump to well-defined entry points in other compartments.

To enforce the access and transition policies, HAKC assigns different MTE colors to the cliques within a
compartment. Importantly, MTE colors are \emph{not} stored directly
in pointers nor are MTE instructions used to check data accesses or control
transfers. Instead, HAKC signs each pointer using PAC, but concatenates as part
of the PAC context a unique compartment identifier and the colors of accessible
cliques within the compartment. This prevents a clique from accessing cliques in
other compartments even though colors may be reused. However, control transfers
to a new compartment must also transfer any data needed from the old
compartment, introducing additional overhead. Since there is no limit to the
number of compartments, HAKC's two-level design thus introduces a trade-off
whereby adding more compartments improves security through finer-grained
compartmentalisation, at the cost of increased overhead caused by data transfers between compartments.

\subsection{Optimizations in Cross-compartment Communication}
\label{sec:mechanism:optimizations}
Apart from a few SFI approaches (\eg \cite{castro2009fast}),
\emph{cross-compartment communication} constitutes the main source of overhead
after compartments are
created~\cite{narayanan2020lightweight,nikolaev_virtuos_2013,narayan2021swivel,Narayanan2019,proskurin2020xmp}.
Many of the kernel compartmentalization mechanisms
in this section
use a range of techniques
to speed up or
mask these costs.
While many of the optimizations
were proposed for specific mechanisms,
their applicability can be considered more widely.

\mypar{Asynchronous Communication}
VirtuOS~\cite{nikolaev_virtuos_2013},
LXDs~\cite{Narayanan2019},
and LVD~\cite{narayanan2020lightweight}
employ asynchronous communication between compartments
to mask the cost of domain switches.
This transforms the programming model from a blocking to a non-blocking one,
and thus some re-engineering of isolated compartments might be required to
perform useful work while waiting for an asynchronous call to complete. To
avoid the engineering effort, an alternative approach is to retain the
blocking programming model but use a cooperative threading library to multiplex
multiple lightweight threads onto a single hardware thread. When a lightweight
thread makes a cross-domain call, it transparently yields to another lightweight
thread in the same domain. This approach also facilitates batching and
pipelining of cross-domain calls made by different lightweight threads.

\mypar{Leveraging Cache Coherence}
Removing the kernel from the critical path in IPC was first proposed by URPC~\cite{bershad1991user}.
LXDs~\cite{Narayanan2019} take this idea one step further:
Instead of using memory as the point of
coherence for cross-domain invocations,
they use
cache coherence to speed-up communication
(in a fashion similar to Barrelfish~\cite{baumann2009multikernel}).
Switching across domains to transfer information becomes unnecessary,
but it requires switching to an asynchronous programming model (as described above)
and pinning domains that need to communicate to different cores.

\mypar{Sparing the Hypervisor} Switching between address spaces of different virtualization domains requires a VM exit,
which is significantly more expensive than a traditional system call~\cite{williams2018unikernels}.
Intel introduced VMFUNC that allows \emph{intra-domain} switching
between a list of predefined extended page tables (EPTs)
with overhead comparable to a system call.
LVD~\cite{narayanan2020lightweight}
uses VMFUNC to speed up cross-domain invocations.
However, there are security implications to using VMFUNC instead of trapping to the hypervisor.
For instance, unlike the case of VM enter/exit, the entry/exit into
the domain is not happening at predefined points,
as the VMFUNC changes the current address view of the domain and continues to execute the next
instruction.
This means that an attacker inside an isolated domain can potentially
find executable bytes that make up the VMFUNC instruction, and trigger an illegal VMFUNC call.
LVD uses two additional techniques to safeguard against this attack vector:
1) Scanning (and rewriting) the executable code to find arbitrary bytes that can form a VMFUNC instruction,
2) Ensuring that the list of allowed EPTs for the VMFUNC is restricted to only two domains i.e.,
the kernel and the isolated domain.
These techniques are generally applicable (or rather needed) in any implementation that uses VMFUNC to compartmentalize the kernel.
xMP~\cite{proskurin2020xmp} uses Virtualization Exceptions (\#VE) %
to handle memory access violations
instead of trapping to the hypervisor.

Both of these techniques give more capabilities
to an isolated compartment
to reduce the number of calls to the hypervisor;
conceptually similar approaches could be employed
by other compartmentalization mechanisms too.
For example,
Dune~\cite{belay2012dune} elevates the capabilities
of a user process to manage its own page table
so that it does not have to trap to the kernel on page faults.
This technique could be applied to user-space drivers
to reduce the number of boundary crossings
between user space and the kernel.

\section{Emerging Hardware Features}
\label{sec:emerging_hw}

In this section,
we discuss
emerging hardware features
that could be used to improve
the performance of different aspects of
kernel compartmentalization mechanisms described
in~\autoref{sec:mechanisms}.

\subsection{Encryption-based Techniques}
\label{sec:emerging_hw:encryption}

AMD~\cite{amdSEV} and Intel~\cite{intelMKTME}
have proposed new hardware extensions
to encrypt pages belonging to a guest VM.
This encryption prevents the hypervisor
from reading the data pages of the VM.
Just as virtualization has been used to implement sandboxes in
the kernel~\cite{nikolaev_virtuos_2013,Narayanan2019,narayanan2020lightweight},
we speculate that these hardware extensions could be adapted to implement safeboxes
to protect data in VMs from the hypervisor.

\subsection{Hardware Capabilities}
\label{sec:emerging_hw:capabilities}

\begin{figure*}[t]
	\includegraphics[width=\textwidth]{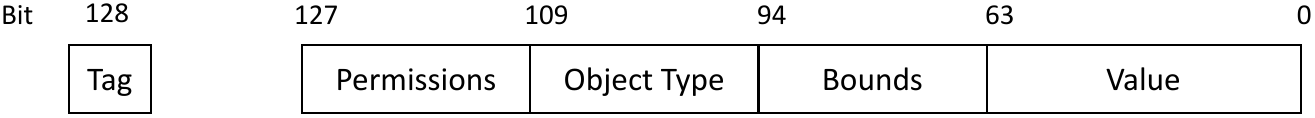}
	\caption{128-bit CHERI capability representation (labelled bit 0--127) with an out-of-band tag (bit 128)~\cite{watson_introduction-to-cheri_nodate}}
	\label{sec:hardware:cheri}
\end{figure*}

CHERI~\cite{watson_cheri_2015} is an extension
to RISC~\cite{woodruff_cheri_2014} and ARM-v8~\cite{morello}
instruction set architectures (ISAs).
It replaces a memory address (pointer)
with a 128-bit \emph{capability} %
and a 1-bit out-of-band \emph{tag}.
As shown in~\autoref{sec:hardware:cheri},
along with the value of an address,
a capability holds additional information about
the memory region being accessed,
e.g., its bounds and
associated permissions.
The capability is considered \emph{valid}
only when the tag bit is set.
When a new capability needs to be created,
it must be derived from a valid source capability.
A legal derivation requires
the bounds and permissions of the derived capability
to be lesser than those of the source capability.
Otherwise,
the hardware clears the new capability's tag bit,
thus making it invalid.
An exception is raised at run time
when an invalid capability is used in a \emph{load},
\emph{store}, or \emph{jump} instruction.
The tag bit cannot be explicitly set for a capability;
therefore,
it is impossible for software to forge capabilities.

When a program accesses a memory location
outside the range specified by the capability (e.g., a buffer overrun),
the hardware raises an exception,
providing \emph{spatial memory safety}.
If the user directly manipulates the address of a jump target,
the corresponding tag bit will be cleared, and
the capability becomes invalid.
Executing an instruction that uses the capability in a jump will
raise an exception, and
thus, CHERI also provides
\emph{control flow integrity}.
Prior work~\cite{sartakov2022capvm} uses CHERI %
to build user-space compartments
to achieve both~properties.

CheriBSD~\cite{watson_cheri_2015} is a port of FreeBSD
to CHERI, in which every pointer in the kernel is a CHERI capability.
At the time of this writing,
CheriBSD does not create compartments inside the kernel using its capabilities.
However, just like Mondrix used Mondrian
for kernel compartmentalization,
CHERI could be used to compartmentalize the kernel in CheriBSD.
There also exists an ongoing effort to port
Linux to the CHERI hardware~\cite{cheri_linux},
but to the best of our knowledge,
similar to CheriBSD,
it does not create compartments
inside the kernel using CHERI.

\subsection{Control Flow Integrity}
\label{sec:emerging_hw:cfi}

Intel Control-Flow Enforcement Technology (CET)~\cite{TechnicalLookIntelCET} provides
two new ISA capabilities to defend against control-flow attacks~\cite{shanbhogue2019security}:
Shadow Stack~\cite{burow2019sok} (SS) and Indirect Branch Tracking~\cite{pappas2013transparent} (IBT).
Shadow Stack (SS) is an additional stack for return addresses only.
The stack has a Shadow Stack Pointer (SSP) pointing to the last address
a program is expected to return to.
If the program returns to a different address, an exception is triggered.
Indirect Branch Tracking (IBT) is a feature that defends against jump/call-oriented programming (JOP and COP) attacks.
By keeping track of the expected targets of indirect branches,
it raises an exception if the target is not one of the expected targets.
CET could be used to optimize software-based
approaches for CFI mentioned in~\autoref{sec:mechanism:intrakerneldomain:sw}
such as KCoFI~\cite{criswell2014kcofi,criswell_secure_2007}.

\section{Evaluating Kernel \Compzz}
\label{sec:evaluation}
\begin{table*}[]
	\centering
	\caption{An overview of the possible factors that may affect the comparability
	of systems. For each selected system, we identify its experimental setup, the
	availability of its source code and documentation, and its reliance on
	hardware-specific features. The last column indicates whether a system
	was compared with similar work.}
	\label{tab:comparibility}
	\resizebox{\textwidth}{!}{%
	\begin{tabular}{|l|l||m{0.16\linewidth}|m{0.14\linewidth}||>{\centering\arraybackslash}m{0.05\linewidth}|c|>{\centering\arraybackslash}m{0.08\linewidth}||c|}
		\cline{1-8}
		\multicolumn{2}{|c||}{System} &
		\multicolumn{2}{c||}{Experimental Setup} &
		\multicolumn{3}{c||}{Reproducibility} &
		\multirow{2}{*}{
			\begin{minipage}[c]{0.09\linewidth} \centering %
				\bigskip{}
				Comparison w/ Other Systems
			\end{minipage}
		} \\ \cline{1-7}
		\multicolumn{1}{|c|}{Name} &
		\multicolumn{1}{c||}{Venue} &
		\multicolumn{1}{c|}{Processor} &
		\multicolumn{1}{c||}{OS} &
		Open Source &
		Documentation &
		Reliance on Hardware-Specific Features &
		\\ \cline{1-8}
		Nooks~\cite{swift2003improving} &
		SOSP '03 &
		Intel Pentium IV &
		Linux 2.4.18 &
		\checkmark &
		\checkmark &
		&
		\\ \cline{1-8}
		Mondrix~\cite{witchel2005mondrix} &
		SOSP '05 &
		Simulator (SimICS/Bochs x86) &
		Linux 2.4.19 &
		&
		&
		\checkmark &
		\\ \cline{1-8}
		SafeDrive~\cite{safedrive06} &
		OSDI '06 &
		Dual Xeon &
		Linux  2.6.15.5 &
		&
		&
		&
		\\ \cline{1-8}
		XFI~\cite{erlingsson2006xfi} &
		OSDI '06 &
		Intel Pentium M &
		Windows (version unknown) &
		&
		&
		&
		\checkmark \\ \cline{1-8}
		Microdrivers~\cite{ganapathy_microdrivers_2008} &
		ASPLOS '08 &
		Intel Pentium D / AMD Opteron &
		Linux 2.6.18.1 &
		&
		&
		&
		\\ \cline{1-8}
		Decaf~\cite{renzelmanndecaf} &
		ATC '09 &
		Intel Pentium D / Intel Core2 Quad &
		Linux 2.6.18.1 &
		&
		&
		&
		\checkmark \\ \cline{1-8}
		BGI~\cite{castro2009fast} &
		SOSP '09 &
		Intel Core2 Duo &
		Windows Vista Enterprise SP1 &
		&
		&
		&
		\\ \cline{1-8}
		LXFI~\cite{mao_software_2011} &
		SOSP '11 &
		Intel i3-550 &
		Linux 2.6.36 &
		&
		&
		&
		\\ \cline{1-8}
		VirtuOS~\cite{nikolaev_virtuos_2013} &
		SOSP '13 &
		Intel Xeon E5520 &
		Linux 3.2.30 &
		\checkmark &
		\checkmark &
		&
		\\ \cline{1-8}
		KCoFI~\cite{criswell2014kcofi} &
		S\&P '14 &
		Intel i7-3770 &
		FreeBSD 9.0 &
		\checkmark &
		\checkmark &
		&
		\\ \cline{1-8}
		PerspicuOS~\cite{dautenhahn2015nested} &
		ASPLOS '15 &
		Intel i7-3770 &
		FreeBSD 9.0 &
		\checkmark &
		\checkmark &
		&
		\\ \cline{1-8}
		KENALI~\cite{song2016enforcing} &
		NDSS '16 &
		Nvidia Tegra K1 &
		Android 5.0-5.1.1 (exact version unclear) &
		\checkmark &
		\checkmark &
		&
		\checkmark \\ \cline{1-8}
		SKEE~\cite{azab_skee_2016} &
		NDSS '16 &
		Qualcomm Snapdragon APQ8084 / Exynos 7420 &
		Linux 3.10.61 (Android 5.1.1)&
		&
		&
		&
		\\ \cline{1-8}
		LXDs~\cite{Narayanan2019} &
		ATC '19 &
		Intel E5-4620 &
		Linux 4.8.4 &
		\checkmark &
		\checkmark &
		\checkmark & %
		\\ \cline{1-8}
		LVD~\cite{narayanan2020lightweight} &
		VEE '20 &
		Intel E5-2660 &
		Linux 4.8.4 &
		\checkmark &
		\checkmark &
		\checkmark & %
		\\ \cline{1-8}
		xMP~\cite{proskurin2020xmp} &
		S\&P '20 &
		Intel i7-7700 &
		Linux 4.18.x &
		\checkmark &
		\checkmark &
		\checkmark & %
		\checkmark \\ \cline{1-8}
		IskiOS~\cite{iskios_raid_21} &
		RAID '21 &
		Intel Xeon Silver 4114 (Skylake) &
		Linux 5.10.x &
		\checkmark &
		&
		\checkmark &
		\\ \cline{1-8}
		HAKC~\cite{mckee2022preventing} &
		NDSS '22 &
		Simulator (ARMv8.5a FAST model) &
		Linux 5.10.x &
		\checkmark &
		\checkmark &
		\checkmark &
		\\ \cline{1-8}
	\end{tabular}
	}
\end{table*}

In this section,
we describe three categories of evaluation criteria used
in current kernel compartmentalization literature,
\ie \emph{performance}, \emph{security}, and \emph{practicality}.
We discuss the difficulties in
comparing among different compartmentalization techniques,
motivating the need for our community
to agree on \emph{standardized benchmarks}
for meaningful comparison.
\autoref{tab:comparibility}
summarizes the experimental setups
and three main factors (\ie availability,
documentation, and reliance on specific hardware features)
that could affect comparability and reproducibility
of the systems we survey. %

\subsection{Performance Evaluation}
\label{sec:eval:performance}
Using only a single performance measure
is insufficient to fully evaluate
a compartmentalized kernel,
because different techniques
can affect the performance of
different aspects of the system.
As such,
existing work uses various \emph{micro-} and \emph{macro-benchmarks}
to measure and understand performance overhead.

\mypar{Microbenchmarks}
The most commonly used microbenchmark
to evaluate core kernel functionality
is LMBench~\cite{mcvoy1996lmbench}, which
measures the performance impact
on individual system calls~\cite{criswell2014kcofi,dautenhahn2015nested,song2016enforcing}.
To measure the overhead on
network throughput,
systems that isolate network drivers~\cite{swift2003improving, mao_software_2011, Narayanan2019, narayanan2020lightweight}
typically use \texttt{netperf}~\cite{netperf} and \texttt{iperf}~\cite{iperf}
to evaluate the overall send/receive bandwidth.
KENALI~\cite{song2016enforcing} and SKEE~\cite{azab_skee_2016}
use the ARM cycle count register (CCNT)
to measure the latency of a round-trip context switch
as the cost of crossing isolated compartments.
SFI-based systems~\cite{mao_software_2011,erlingsson2006xfi}
use the SFI microbenchmarks~\cite{small1998misfit}
to measure the overhead on tasks
performed by a wide range of kernel extensions,
such as page eviction, MD5 fingerprinting, and logical disk operations.

\mypar{Macrobenchmarks}
In contrast to microbenchmarks,
macrobenchmarks can provide good indicators
of a system's performance under realistic workloads.
Network applications
like Apache web servers
make heavy use of kernel functionality,~so existing systems~\cite{criswell_memory_2009, criswell2014kcofi,
dautenhahn2015nested, swift2003improving}
use ApacheBench~\cite{apachebench} and httperf~\cite{mosberger1998httperf}
to measure the overhead on a web server's bandwidth.
Besides network servers,
some systems also perform macrobenchmarking
using client-side applications.
For example,
SVA~\cite{criswell_memory_2009}
measures file conversion time (from WAV to MP3)
using the LAME MP3 encoding benchmark~\cite{lame_mp3}
and file compression time
of \texttt{bzip2}
to evaluate its performance overhead
on end-user applications.
KENALI~\cite{song2016enforcing} and SKEE~\cite{azab_skee_2016} use
Android benchmarks,
such as AnTuTu~\cite{antutu},
Geekbench~\cite{geekbench},
and Vellamo~\cite{vellamo},
to simulate user activities like web browsing and gaming
on their Android compartmentalization prototypes.

\subsection{Security Evaluation}

Security evaluation is used to
verify the effectiveness of
a kernel compartmentalization mechanism
against faults
and vulnerabilities.
To evaluate fault-isolating and fault-resistant
systems
(see \autoref{sec:categorisation:exclave}),
\emph{fault injection} is used 
to validate that
they isolate the kernel
from faults
that would otherwise lead to undefined behavior
or system failure.
For example,
Nooks~\cite{swift2003improving} adapted a synthetic fault injector
to emulate programming errors
(\eg uninitialized local variables)
in kernel extensions.

To evaluate a system's ability
to mitigate vulnerabilities, authors either perform experiments with real-world exploits from \emph{Common Vulnerabilities and Exposures} (CVEs) (a database of publicly disclosed cyber-threats)~\cite{swift2003improving,witchel2005mondrix,ganapathy_microdrivers_2008,castro2009fast,criswell_memory_2009,mao_software_2011,nikolaev_virtuos_2013,song2016enforcing,proskurin2020xmp}, 
or provide case studies of a vulnerability or a class of vulnerabilities~\cite{iskios_raid_21,mckee2022preventing}. 
The former replicates CVEs that match the threat model assumptions, evaluating a kernel compartmentalization system against the selected real-world exploits to test its efficacy.
For example,
LXFI~\cite{mao_software_2011}
evaluates its ability to thwart privilege escalation vulnerabilities by transmitting crafted messages over reliable datagram sockets, triggering an exploit that allows an attacker to execute arbitrary code in the kernel~\cite{cve2010_3904}. 
The latter focuses on analyzing how an attack \emph{would} be mitigated by a kernel compartmentalization system, assuming that the system correctly implements its design specifications.

Quantitative metrics also exist
to measure the extent to which
a compartmentalization mechanism reduces the attack surface.
For example,
KCoFI~\cite{criswell2014kcofi} uses
the \emph{average indirect target reduction} (AIR) metric~\cite{zhang2013control}
to quantify the reduction
of possible targets
of indirect control-flow transfer,
as well as the open-source ROPgadget tool~\cite{ropgadgettool}
to measure the reduction of possible ROP gadgets.
KCoFI and PerspicuOS~\cite{dautenhahn2015nested}
both use the SLOCCount tool~\cite{sloccount}
to measure a kernel's TCB size (in lines of code)
as an indirect measure of its attack surface.

\subsection{Practicality Evaluation}
\label{sec:evaluation:effort}

Evaluating the practicality of
a kernel compartmentalization mechanism
through the amount of \emph{engineering effort}
helps identify potential barriers
to its adoption.
One measure of
such an effort
is the number of lines of code that are modified or added.
For example,
PerspicuOS~\cite{dautenhahn2015nested}
counts code modifications in Git change logs
to quantify the effort
of porting a commodity kernel
to the nested kernel architecture.
LXFI~\cite{mao_software_2011} and Microdriver~\cite{ganapathy_microdrivers_2008}
use the number of annotations
to measure the engineering overhead
of modifying kernel modules.

\subsection{Limitations of current evaluations}

It is difficult %
to understand how a given kernel compartmentalization approach compares to another,
because
different techniques have been implemented over a multitude of OSs.
For example, 
BGI~\cite{castro2009fast} and LXFI~\cite{mao_software_2011},
while similar in many aspects,
are implemented in Windows and Linux, respectively.
Different hardware used in evaluation also makes direct comparison difficult.
For example, Mondrix~\cite{witchel2005mondrix} requires
hardware that supports Mondrian Memory Protection,
which is not widely available.
Even when two approaches use the same OS,
their performance can still vary significantly
across
OS \emph{versions} or 
as a result of misconfigurations~\cite{ren2019analysis}.
In addition, two solutions may assume
slightly different threat models,
which lead to different sets of trades-offs (see \autoref{sec:categorisation}).

\noindgras{Performance. }We suggest that
future work evaluate the overhead
introduced by common compartmentalization operations, %
such as the transitioning cost,
against other systems
(or at a minimum, against some well-known ``cost''
such as function calls or system calls)
to facilitate order-of-magnitude comparison.
Unfortunately,
only a few evaluations provide such order-of-magnitude comparisons~\cite{Narayanan2019,narayanan2020lightweight}.
We acknowledge the difficulty of comparing different solutions.
However, some baselines, 
such as the SFI microbenchmark~\cite{small1998misfit}
frequently used to evaluate SFI-based kernel compartmentalization systems, 
should be adopted to provide a common frame of reference.
For example, 
LXFI compares its SFI microbenchmark results to those of XFI
without the need to re-run the same experiments on XFI.

\noindgras{Security Effectiveness.} Vulnerabilities are hard to replicate
as they get patched out over time.
To evaluate whether two competing systems mitigate the same vulnerability,
forward- or back-porting of these approaches
to the same affected OS (and the same OS version)
would require significant engineering effort,
arguably beyond reasonable evaluation expectations.
As a consequence, the set of CVEs used to perform evaluation 
is inconsistent, 
making it difficult to generalize from a specific CVE
(or a class of CVEs)
the effectiveness of an approach.
Many approaches also have different threat models,
which further complicates comparison 
(\eg some types of vulnerabilities are out of scope for one system but in scope for others).
We suggest that future work adopt
standardized approaches
to quantitatively measure the reduction of the attack surface~\cite{zhang2013control,ropgadgettool} and the TCB~\cite{sloccount}.
While not a panacea, it provides a common frame of reference
without an unreasonable amount of labor.

\noindgras{Engineering Effort.} Only a few authors have attempted to
\emph{quantitatively} evaluate %
the effort required to isolate a kernel component 
e.g., by counting the number of manual annotations~\cite{dautenhahn2015nested,mao_software_2011,ganapathy_microdrivers_2008,renzelmanndecaf} (see \autoref{sec:evaluation:effort}).
These measurements can be comparable between two competing systems, 
if they share the \emph{same} isolated kernel component.
For example, 
LXFI~\cite{mao_software_2011} and Decaf~\cite{renzelmanndecaf} both measured the number of annotations 
on the E1000 network device driver.
Qualitative evaluation
is more common, however.
For example,
SKEE~\cite{azab_skee_2016} only described
the effort needed to modify the kernel,
without any data provided to quantify that effort.
While a qualitative description is useful,
it provides only a \emph{subjective}
comparison across systems.
We suggest that future work at a minimum quantitatively measure
the amount of code that needs to be written
and the number of files/subsystems modified.

It is equally important
to understand
the difficulty
of \emph{identifying} compartment boundaries,
which is different from
the complexity of the implementation code
that enforces those boundaries.
We observe that
measuring the intellectual load required 
to identify appropriate compartment boundaries 
or to provide hints (and annotations) 
to an automated system 
remains relatively unexplored.
However,
this is of particular importance 
when deciding if a given approach is practical
to adopt (\autoref{sec:boundaries}).

\section{Conclusion}
\label{sec:conclusion}
We survey operating system kernel compartmentalization work
spanning more than two decades.
We identify two main categories of compartmentalization:
\emph{sandbox} systems that protect the kernel 
from untrusted %
kernel extensions 
and \emph{safebox} systems 
that isolate security-sensitive mechanisms %
from the rest of the kernel.
For each type of compartmentalization,
we systematize the techniques used 
to identify boundaries between security domains 
and enforce isolation across those boundaries.
We describe various hardware features 
that can be adopted to improve
the performance of kernel compartmentalization
and thus facilitate its practical adoption.
Finally,
we identify barriers
that lead to inconsistent evaluation strategies
among different solutions
and suggest ways to address them.

\section*{Acknowledgment}
We acknowledge the support of the Natural Sciences and Engineering Research Council of Canada (NSERC). Nous remercions le Conseil de
recherches en sciences naturelles et en génie du Canada (CRSNG) de son soutien.

\bibliographystyle{ACM-Reference-Format}
\bibliography{biblio}

\end{document}